\documentclass{aa}

\usepackage[switch,columnwise]{lineno}
\usepackage{natbib,url,twoopt}
\usepackage[varg]{txfonts}           
\usepackage[hyperindex,pdftex]{hyperref}
\usepackage{pdfcomment}              
\usepackage{acronym}                 
\usepackage{cleveref}                  
\usepackage{here}
\usepackage{bm}
\hypersetup{
 colorlinks=true,%
 linkcolor=cyan,
 citecolor=cyan,
}

\bibpunct{(}{)}{;}{a}{}{,}    

\makeatletter
\newcommand{\bibnote}[2]{\global\@namedef{#1note}{#2}}
\newcommand{\biblink}[2]{\global\@namedef{#1link}{#2}}
\newcommand{\Tabref}[1]{Table~\ref{#1}}
\newcommand{\Equref}[1]{Eq.~(\ref{#1})}
\newcommand{\Figref}[1]{Fig.~\ref{#1}}

\crefname{figure}{Fig.}{Figs.}
\crefname{equation}{Eq.}{Eqs.}
\makeatother

\makeatletter
 \newcommandtwoopt{\citeads}[3][][]{%
   \nonstopmode
   \href{http://adsabs.harvard.edu/abs/#3}%
        {\def\hyper@linkstart##1##2{}%
         \let\hyper@linkend\@empty\citealp[#1][#2]{#3}}
   \biblink{#3}{\href{http://adsabs.harvard.edu/abs/#3}{ADS}}%
   \errorstopmode}            
 \newcommandtwoopt{\citepads}[3][][]{%
   \nonstopmode
   \href{http://adsabs.harvard.edu/abs/#3}%
        {\def\hyper@linkstart##1##2{}%
         \let\hyper@linkend\@empty\citep[#1][#2]{#3}}
   \biblink{#3}{\href{http://adsabs.harvard.edu/abs/#3}{ADS}}%
   \errorstopmode}            
 \newcommandtwoopt{\citetads}[3][][]{%
   \nonstopmode
   \href{http://adsabs.harvard.edu/abs/#3}%
        {\def\hyper@linkstart##1##2{}%
         \let\hyper@linkend\@empty\citet[#1][#2]{#3}}
   \biblink{#3}{\href{http://adsabs.harvard.edu/abs/#3}{ADS}}%
   \errorstopmode}            
 \newcommandtwoopt{\citeyearads}[3][][]{%
   \nonstopmode
   \href{http://adsabs.harvard.edu/abs/#3}%
        {\def\hyper@linkstart##1##2{}%
         \let\hyper@linkend\@empty\citeyear[#1][#2]{#3}}
   \biblink{#3}{\href{http://adsabs.harvard.edu/abs/#3}{ADS}}%
   \errorstopmode}            
\makeatother

\newacro{ADS}{Astrophysics Data System}
\newacro{NLTE}{non-local thermodynamic equilibrium}
\newacro{NASA}{National Aeronautics and Space Administration}

\begin{document}

   \title{Gas flow around a planet embedded in a protoplanetary disc:}

   \subtitle{the dependence on the planetary mass}

          \author{Ayumu Kuwahara\inst{1}
          \thanks{\email{kuwahara.a.aa@m.titech.ac.jp}} 
          \and Hiroyuki Kurokawa\inst{2}
          \and Shigeru Ida\inst{2}}

   \institute{Department of Earth and Planetary Sciences, Tokyo Institute of Technology, Ookayama, Meguro-ku, Tokyo, 152-8551, Japan
         \and
             Earth-Life Science Institute, Tokyo Institute of Technology, Ookayama, Meguro-ku, Tokyo, 152-8550, Japan}

   \date{Received September XXX; accepted YYY}

 
  \abstract
   {The ubiquity of short-period super-Earths remains a mystery in planet formation, as these planets are expected to become gas giants via runaway gas accretion within the lifetime of a protoplanetary disc. Super-Earths' cores should form in the late stage of the disc evolution to avoid runaway gas accretion.}
   {The three-dimensional structure of the gas flow around a planet is thought to influence the accretion of both gas and solid materials. In particular, the outflow in the mid-plane region may prevent the accretion of the solid materials and delay the formation of super-Earths' cores. However, it is not yet understood how the nature of the flow field and outflow speed change as a function of the planetary mass. In this study, we investigate the dependence of gas flow around a planet embedded in a protoplanetary disc on the planetary mass.}
   {Assuming an isothermal, inviscid gas disc, we perform three-dimensional hydrodynamical simulations on the spherical polar grid, which has a planet located at its centre.}
   {We find that gas enters the Bondi or Hill sphere at high latitudes and exits through the mid-plane region of the disc regardless of the assumed dimensionless planetary mass $m=R_{\rm Bondi}/H$, where $R_{\rm Bondi}$ and $H$ are the Bondi radius of the planet and disc scale height, respectively. The altitude from where gas predominantly enters the envelope varies with the planetary mass. The outflow speed can be expressed as $|u_{\rm out}|=\sqrt{3/2}mc_{\rm s}$ $(R_{\rm Bondi}\leq R_{\rm Hill})$ or $|u_{\rm out}|=\sqrt{3/2}(m/3)^{1/3} c_{\rm s}$ ($R_{\rm Bondi}\geq R_{\rm Hill}$), where $c_{\rm s}$ is the isothermal sound speed and $R_{\rm Hill}$ is the Hill radius. The outflow around a planet may reduce the accretion of dust and pebbles onto the planet when $m\gtrsim\sqrt{\rm St}$, where St is the Stokes number.}
   {Our results suggest that the flow around proto-cores of super-Earths may delay their growth and, consequently, help them to avoid runaway gas accretion within the lifetime of the gas disc.}


   \maketitle
%

\section{Introduction}     \label{sec:introduction}
The Kepler mission has found that about $\sim$50\% of Sun-like stars harbor short-period super-Earths with  orbital periods less than 85 days and radii of 1--4 $R_{\oplus}$ (Earth radius) \cite[e.g.,][]{Fressin:2013}. Radial velocity measurements and transit timing variations have also revealed that the masses of those planets are in the range of 2--20 $M_{\oplus}$ (Earth mass) \cite[e.g.,][]{Weiss:2014}. The reason for the ubiquity of short-period super-Earths has not been fully elucidated by planet formation theory.

According to the core accretion model, when the total mass of a planet has reached critical core mass, $M_{\rm crit}\sim10$ $M_{\oplus}$, runaway gas accretion is triggered and it evolves into a gas giant \cite[e.g.,][]{Mizuno:1980,Pollack:1996,Ikoma:2000}. The runaway time-scale is about 1 Myr for a solid core having 10 $M_{\oplus}$, which is comparable to the typical disc lifetime, $\sim$ a few Myr, and the time-scale is much shorter when the atmosphere is dust free \cite[]{Lee:2014}. Short-period super-Earths have avoided runaway gas accretion and growth into gas giants within the lifetime of the disc.

Hydrodynamic effects in a disc have been proposed as one solution to avoid runaway gas accretion \cite[]{Ormel:2015b}. Protoplanetary disc gas enters the Bondi sphere of a planet embedded in a disc at high latitudes and leaves it through the mid-plane regions. They have argued that the continuous recycling of atmosphere within the Bondi sphere is faster than the cooling of the envelope gas, and so that further accretion of disc gas is prevented; though the efficiency of the atmospheric recycling is a controversial issue \cite[]{Cimerman:2017,Lambrechts:2017,Kurokawa:2018}. In addition, the dominance of disc-wind-driven accretion over viscous accretion onto the star may induce a supply limit to the gas accretion onto super-Earths' cores \cite[]{OgiharaHori:2018}.

The late-stage core formation model has also been suggested as another solution. In this scenario, super-Earths' cores are assumed to have formed via coagulation of proto-cores during disc dispersal \cite[]{Lee:2014}. The final assembly during disc dispersal is the expected result in conventional planet formation theory due to mutual gravitational interactions of proto-cores \cite[e.g.,][]{Kominami:2002,Inamdar:2015}. The gas accretion during the limited period before the disc dispersal results in super-Earths having envelopes $\sim$1--10\% their mass \cite[]{Ikoma:2012,Owen:2015,Ginzburg:2016}. 

The plausible late-stage core formation scenarios involve the migration of super-Earths' cores formed at distant orbits \cite[]{Tanaka:2002,Ogihara:2009,Ida:2010}. After forming proto-cores beyond the snow line, they begin to migrate inwards. Super-Earths form in the inner region of the disc via giant impact. The idea is supported by the inference that some of the low-density super-Earths may contain large amounts of water \cite[]{Leger:2004,Selsis:2007,Rogers:2010,Valencia:2010,Lopez:2013,Weiss:2014}. The presence of water on the super-Earths has also been suggested observationally. From the observations of some super-Earths---for instance, GJ 1214b---a featureless transmission spectrum has been found in the observed band, which suggests that the planet's atmosphere could be dominated by relatively heavy molecules, such as water, or it could contain extensive high-altitude clouds or haze \cite[]{Narita:2013,Kreidberg:2014}. 

The feasibility of the late-stage core formation scenario depends on the planet formation regimes. 

In the pebble accretion theory \cite[]{Ormel:2010,Lambrechts:2012}, cores accrete particles with radii of approximately mm--cm drifting from the outer region of the disc. In this scenario, pebble isolation mass becomes large in the outer region of the disc, $M_{\rm iso}^{\rm pebble}\sim10$ $M_{\oplus}$ at 1 au \cite[]{Lambrechts:2014}. Proto-cores can become rather heavy, which leads to runaway gas accretion and the evolution of cores into gas giants. Therefore, it is necessary to suppress pebble accretion.

The horseshoe flows extended in the anterior-posterior direction in the planet's orbital direction have a characteristic vertical structure like a column, and a fraction of the horseshoe flow sharply descends towards the planet due to the planet's gravity \cite[]{Fung:2015}. They have reported that outflow from the Bondi sphere at the mid-plane region has the speed of the order of isothermal sound speed, $\sim c_{\rm s}$. This outflow has the potential to affect the accretion of solid materials to the core of the planet and may delay its growth. The flow field around the planet affects the accretion rate of solid materials. In 2D simulations, trajectories of small solid particles varies with conditions, and accretion of these particles may be suppressed in the case of small dust-size particles  \cite[]{Ormel:2013}. In the 3D case, small particles (10 $\mu$m--1 cm) move away from the planet in the horseshoe flow \cite[]{Popovas:2018}. 

Whereas there are many previous studies with different calculation settings (isothermal or non-isothermal, inviscid or viscous, local or global frame, etc), it is unclear how the nature of the flow field changes as a function of the mass of the planet. In this study, therefore, we focus on the dependence of the flow field on the planetary mass and principally investigate the speed of outflow.

The structure of this paper is as follows. In Section \ref{sec:method} we describe the numerical method. In Section \ref{sec:result} we show the results obtained from a series of simulations and present an analytic estimate of outflow speed. In Section \ref{sec:discussion} we discuss the implications for the formation of super-Earths. We summarise in Section  \ref{sec:conclusion}.
\begin{table*}[!tp]
\caption{Lists of the simulations. The following columns give the simulation name, the size of the Bondi radius of the planet, the size of the Hill radius of the planet, the size of the outer edge of the calculation domain, the length of the injection time, the length of the calculation time, and the resolution, respectively. Each run has the resolution; $[\log r,\theta,\phi]=[128,64,128],\ [100,50,100]$, and $[160,80,160]$ for the fiducial, low, and high resolution simulations, respectively.}
\centering
\begin{tabular}{ccccccc}\hline
Name & $R_{\rm Bondi}\ [H]$ & $R_{\rm Hill}\  [H]$ & $r_{\rm out}\ [H]$ & $t_{\rm inj}\ [\Omega^{-1}]$ & $t_{\rm end}\ [\Omega^{-1}]$ & Resolution\\ \hline\hline
\texttt{m001}, \texttt{m001-low, m001-high} & 0.01 & 0.15 & 0.5 &  0.5 & 10 & fiducial, low, high\\
\texttt{m005}, \texttt{m005-low, m005-high} & 0.05 & 0.26 & 0.5 &  0.5 & 30 & fiducial, low, high \\
\texttt{m01}, \texttt{m01-low, m01-high} & 0.1 & 0.32 & 0.5 &  0.5 & 50 & fiducial, low, high \\
\texttt{m05}, \texttt{m05-low} & 0.5 & 0.55 & 5.0 &  1.0 & 100 & fiducial, low \\
\texttt{m1}, \texttt{m1-low} & 1.0 & 0.69 & 5.0 &  1.0 & 100 & fiducial, low \\
\texttt{m2}, \texttt{m2-low} & 2.0 & 0.87 & 10.0 &  1.0 & 100 & fiducial, low \\ 
\texttt{m001-extendD} & 0.01 & 0.15 & 1.0 &  0.5 & 10 & fiducial \\
\texttt{m005-extendD} & 0.05 & 0.26 & 1.0 & 0.5 & 30 & fiducial \\
\texttt{m01-extendD} & 0.1 & 0.32 & 1.0 & 0.5 & 50 & fiducial \\ \hline
\end{tabular}
\label{tab:simulation}
\end{table*}

\section{Methods}    \label{sec:method}
In this study, we performed three-dimensional hydrodynamical simulations of protoplanetary disc gas around a planet, and investigated how the nature of the flow field changes as a function of the planetary mass. Most of our methods of the simulations followed that of \cite{Kurokawa:2018}. Though they have focused on the differences between isothermal and non-isothermal simulations, we focused on the dependence of the flow field on the planetary mass and conducted detailed studies.
\subsection{Dimensionless units}
The scale of the lengths, times, velocities, and densities are normalised by disc scale height $H$, the reciprocal of the orbital frequency $\Omega^{-1}$, the isothermal sound speed $c_{\rm s}$, and the gas density at planetary orbit $\rho_{\rm disc}$, respectively. In this dimensionless unit system, the dimensionless mass of the planet is expressed by the ratio of the Bondi radius of the planet, $R_{\rm Bondi}$, to the scale height of the disc,
\begin{align}
m\equiv\frac{R_{\rm Bondi}}{H}=\frac{GM_{\rm p}}{c_{\rm s}^{3}/\Omega},
\end{align}
where $G$ is the gravitational constant, and $M_{\rm p}$ is the mass of the planet. When we assume a solar-mass star and a disc temperature profile $T=270\left(a/1\ \text{au}\right)^{-1/2}$ K, which corresponds to the minimum-mass solar nebula model \cite[]{Weidenschilling:1977,Hayashi:1985}, $M_{\rm p}$ is described by
\begin{align}
M_{\rm p}\simeq12m\left(\frac{a}{1\ \text{au}}\right)^{3/4}M_{\oplus},\label{eq:planetarymass}
\end{align}
where $a$ is the orbital radius \cite[]{Kurokawa:2018}. The dimensionless planetary mass $m=0.01$ corresponds to a planet of 0.12 $M_{\oplus}$ revolving around a solar-mass star at 1 au (\Equref{eq:planetarymass}). 

Under this dimensionless unit, the Hill radius of the planet is given by 
\begin{align}
R_{\rm Hill}=\left(\frac{m}{3}\right)^{1/3}H,\label{eq:hillradius}
\end{align}
and the following relationships, $R_{\rm Bondi}\leq R_{\rm Hill}$ and $R_{\rm Hill}\leq H$ hold when $m\leq0.58$ and $m\leq3$, respectively.

\subsection{Governing equations}
The dimensionless governing equations for isothermal and inviscid fluid in the dimensionless unit are described as below.
\begin{align}
&\frac{\partial \rho}{\partial t}+\nabla\cdot\rho{\bm v}=0,\label{eq:continuity}\\
&\left(\frac{\partial }{\partial t}+\bm{v}\cdot\nabla\right)\bm{v}=-\frac{1}{\rho}\nabla P+\sum_{i}\bm{F}_{i},\label{eq:euler} \\
&P=\rho\label{eq:state},
\end{align} 
where $\rho$ is the density, $t$ is the time, $\bm{v}$ is the velocity, $P$ is the pressure, and $\bm{F}_{i}$ is the external force, respectively. Our simulations were performed on the spherical polar coordinate co-rotating with a planet at the orbital frequency $\bm{\Omega}$. Therefore, the second term in the RHS of \Equref{eq:euler} consists of the following elements: the Coriolis force $\bm{F}_{\rm cor}=-2\bm{\Omega}\times\bm{v}$, the tidal force $\bm{F}_{\rm tid}=3x\Omega^{2}\bm{e}_{x}-z\Omega^{2}\bm{e}_{z}$, and the gravitational force
\begin{align}
\bm{F}_{\rm grav}=-\nabla\Phi_{\rm p}\left\{1-\exp\left[-\dfrac{1}{2}\left(\frac{t}{t_{\rm inj}}\right)^{2}\right]\right\},
\end{align}
where $t_{\rm inj}$ is the injection time and $\Phi_{\rm p}$ is the gravitational potential expressed by 
\begin{align}
&\Phi_{\rm p}=-\frac{m}{\sqrt{r^{2}+r_{\rm s}^{2}}},
\end{align}
where $r$ is the distance from the centre of the planet, and $r_{\rm s}$ is the softening length in the $r$-direction. We set the softening length to be equal to 7\% of the Bondi radius of the planet for all simulations.  \cite{Ormel:2015b} have reported that a rapid increase of the gravity of the planet in the unperturbed disc affects the results of the simulations. To avoid this numerical problem, the planet's gravity is gradually inserted into the disc at the injection time, $t_{\rm inj}$. 

\subsection{Disc model}
A planet is embedded in an isothermal, inviscid gas disc and is orbiting around the central star at the distance $a$ with the orbital frequency $\Omega=\sqrt{GM_{\ast}/a^{3}}$, where $M_{\ast}$ is the mass of the host star. The Bondi radius of the planet, $R_{\rm Bondi}=GM_{\rm p}/c_{\rm s}^{2}$, is assumed to be larger than the physical radius of the planet. In most of our simulations, $R_{\rm Bondi}$ was smaller than the disc scale height, $H$. 

We consider the vertical structure of the density distribution in the disc,
\begin{align}
\rho_{\infty}(z)=\rho_{\rm disc}\exp\left[-\frac{1}{2}\left(\frac{z}{H}\right)^{2}\right],\label{eq:density}
\end{align}
where $\rho_{\rm disc}$ is the density of the unperturbed disc, $z$ was the distance in the vertical direction. The density stratification is assumed to be the initial state of the disc. We set \Equref{eq:density} as the boundary condition at the outer edge of the computational domain (see subsection \ref{sec:boundary}).

Keplerian shear existed in the unperturbed state of the disc and had the velocity
\begin{align}
{\bm v}_{\infty}(x)=-\frac{3}{2}x\Omega\bm{e}_{y},\label{eq:Keplerianshear}
\end{align}
where $\bm{e}_{y}$ is the unit vector in the $y$-direction. We did not include the headwind of the disc (namely, the gas disc rotates Keplerian).

\subsection{Boundary conditions} \label{sec:boundary}
Our three-dimensional hydrodynamical simulations were performed on the spherical polar coordinates ($r$, $\theta$, $\phi$) having the specific boundary conditions as described below.

Since all of our simulations were performed under the inviscid condition, we introduced the free-slip inner boundary at $r=r_{\rm inn}$ to prevent the loss of mass in the radial direction, where $r_{\rm inn}$ is the inner edge of the computational domain. We adopted $r_{\rm inn}=10^{-3}$.

We also introduced the outer edge of the computational domain at $r=r_{\rm out}$ where the density and the velocity have constant values, $\rho=\rho_{\infty}(z)$, and ${\bm v}={\bm v}_{\infty}(x)$. The choice of the domain size will be justified in subsection \ref{sec:result3} and Appendix \ref{sec:appendix}. For the azimuthal direction, a periodic boundary condition was introduced such that the relationship $A(r,\theta,\phi)=A(r,\theta,\phi+2\pi)$ holds for an arbitrary scalar or vector quantity $A$.

We defined the simulations having the resolution $[\log r,\theta,\phi]=[128,64,128]$ as the fiducial models. We also performed low and high resolution simulations having $[\log r,\theta,\phi]=[100,50,100]$ and $[160,80,160]$, respectively. We confirmed that the numerical convergence was achieved: the key results do not depend on the resolution (see Appendix \ref{sec:appendix} for details). We adopted a logarithmic grid for the radial coordinate, which has a higher resolution in the vicinity of the planet.

\subsection{Athena++ code}
To perform our simulations, we used Athena++ code, which was a complete re-write of the Athena astrophysical magnetohydrodynamics (MHD) code (\cite{White:2016}, Stone et al. in prep). Athena++ provides a Python script for reading data from output files of the calculations and, in this study, we performed various analyses by using this script. In some of the results of analysis, we averaged some components of the velocities in the azimuthal direction on the mid-plane of the disc (see section \ref{sec:result}). If a certain component of the velocity and the density of the flow field are given as continuous functions, the weighted average velocity in the azimuthal direction is described as 
\begin{align}
\langle v_{\lambda}\rangle_{\phi}=\dfrac{\int_{0}^{2\pi}\rho_{\rm gas}v_{\lambda}d\phi}{\int_{0}^{2\pi}\rho_{\rm gas} d\phi},
\end{align}
where $\rho_{\rm gas}$ is the gas density and $v_{\lambda}$ ($\lambda=r,\ \theta,\ \phi$) is a certain component of the velocity of gas. In a series of results obtained from our grid simulations by using Athena++ code, any physical quantities were given as discrete data on grid points of each grid divided into $N_{r}\times N_{\theta}\times N_{\phi}=128\times64\times128$, where $N_{r},\ N_{\theta},\ N_{\phi}$ meant the number of the grid sections in each direction. We set it such that each grid number in each direction is expressed as $(r,\ \theta,\ \phi)=(i,\ j,\ k)$, and since $j=32$ corresponds to the mid-plane of the disc, a certain component of the velocity on an arbitrary grid is represented by $v_{\lambda_{i,32,k}}$. Therefore, azimuthally averaged velocity on the $i$th grid was calculated by
\begin{align}
    \langle v_{\lambda_{i,32}}\rangle_{\phi}=\frac{\displaystyle\sum_{k=0}^{127}\rho_{{\rm gas}_{i,32,k}}v_{\lambda_{i,32,k}}}{\displaystyle\sum_{k=0}^{127}\rho_{{\rm gas}_{i,32,k}}},
\end{align}
where $\rho_{{\rm gas}_{i,j,k}}$ and $v_{\lambda_{i,j,k}}$ are the density and the $\lambda$-component of the velocity on the arbitrary coordinate has the grid number ($i,j,k$).
\subsection{Parameter sets}
All of our simulations are listed in \Tabref{tab:simulation}. When the Hill radius of the planet exceeds the size of the disc scale height ($m>3$), a gap forms close to the planet's orbit in a disc \cite[]{Lin:1993}. Our local simulations, however, could not handle the gap opening. For this reason we only handled a range of planetary masses, $m=0.01$--2, in a series of simulations. 

Through performing test simulations several times, we confirmed that an unphysical flow pattern emerged in the vicinity of the planet in the early stage of the time evolution of the flow field when $t_{\rm inj}$ was short, especially for planets with $m\geq0.5$. Therefore, we set the length of the injection time to be longer for the planets with $m\geq0.5$. We also found that it takes a longer time for the flow field to reach the steady state, particularly for the planets with $m\geq0.05$. Accordingly, we set the $t_{\rm end}$ long enough for the flow field to reach the steady state in all of our simulations.

 \begin{figure*}[htbp]
 \resizebox{\hsize}{!}
 {\includegraphics{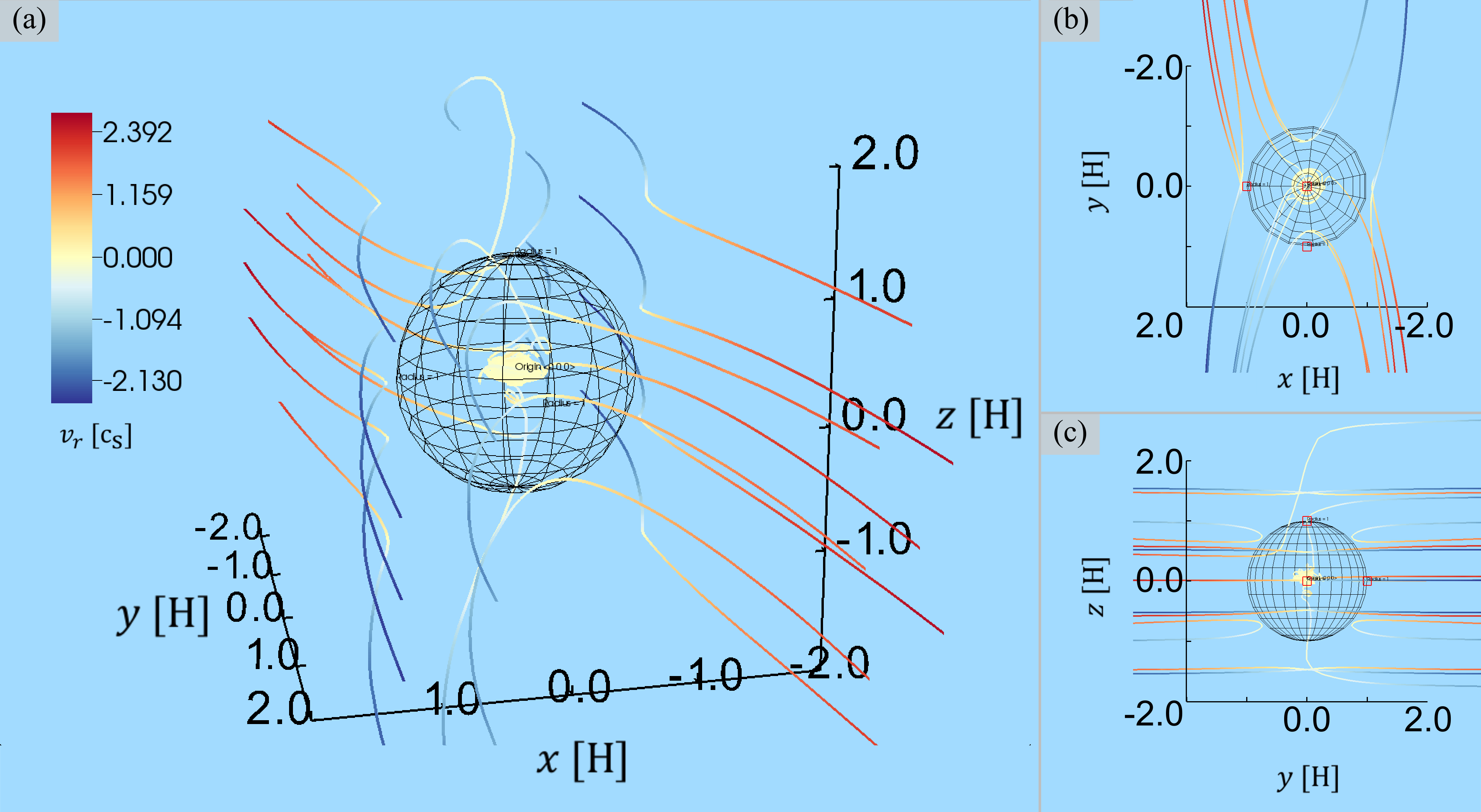}} 
 \caption{The 3D structure of the flow field obtained from \texttt{m1} run at $t=100$. The solid lines and the sphere represent the characteristic streamlines of gas flow around the planet and the Bondi region, respectively. The colour shows the radial velocity normalised by the isothermal sound speed. The region where $v_{r}$ has positive (outflow) and negative value (inflow) are shown in blue and red, respectively. (a): The perspective view of the flow field. (b): The $x$-$y$ plane viewed from $+z$ direction. (c): The $y$-$z$ plane viewed from $-x$ direction.}
\label{fig:3Da}
\end{figure*}

\begin{figure*}[!htbp]
  \centering
    \begin{tabular}{c}
 
 
      \begin{minipage}{0.50\hsize}
        \centering
          \includegraphics[keepaspectratio, width=\linewidth, angle=0]
                          {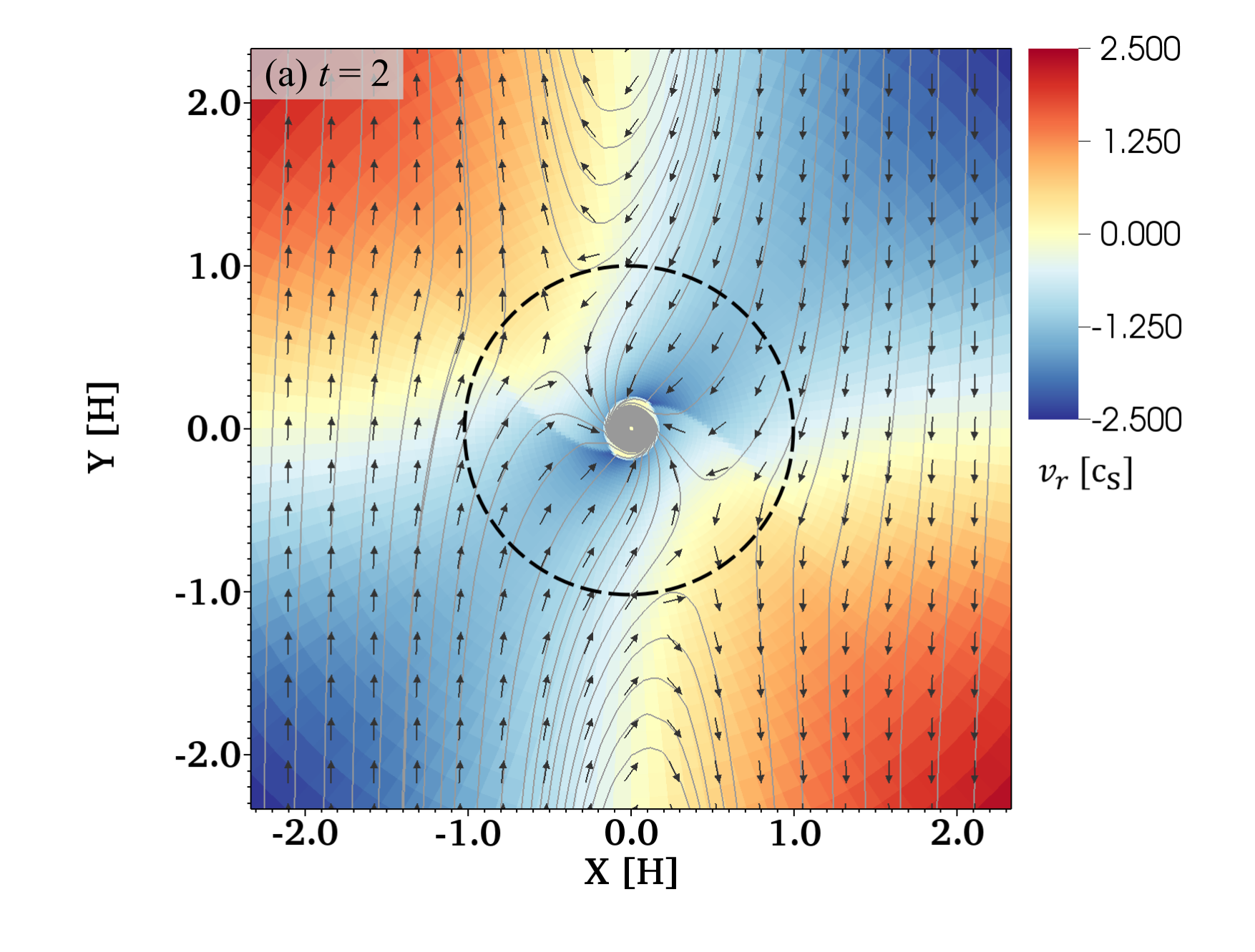}
      \end{minipage}
 
 
      \begin{minipage}{0.50\hsize}
        \centering
          \includegraphics[keepaspectratio, width=\linewidth, angle=0]
                          {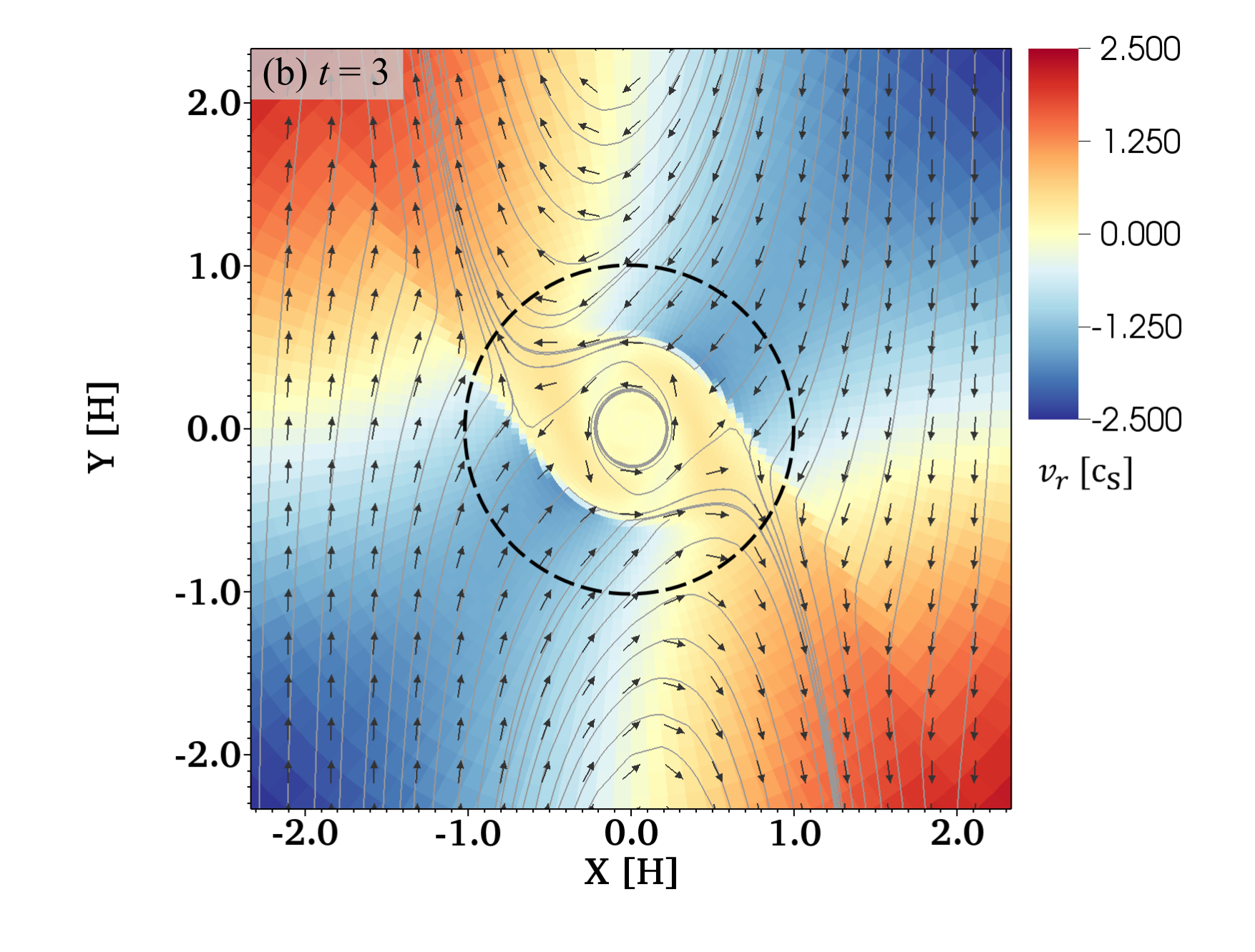}
      \end{minipage} \\
 
 
      \begin{minipage}{0.50\hsize}
        \centering
          \includegraphics[keepaspectratio, width=\linewidth, angle=0]
                          {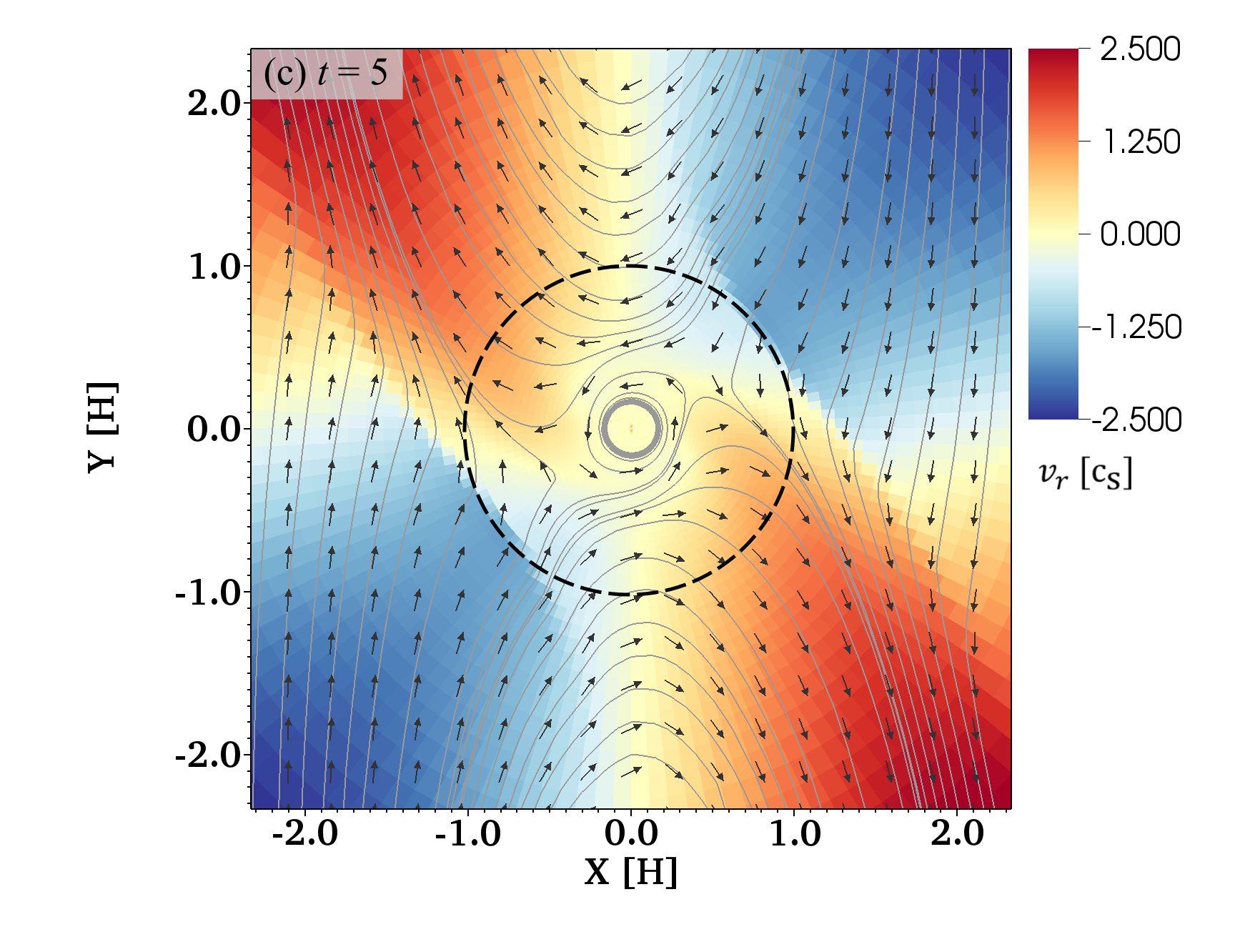}
      \end{minipage}
 
 
      \begin{minipage}{0.50\hsize}
        \centering
          \includegraphics[keepaspectratio, width=\linewidth, angle=0]
                          {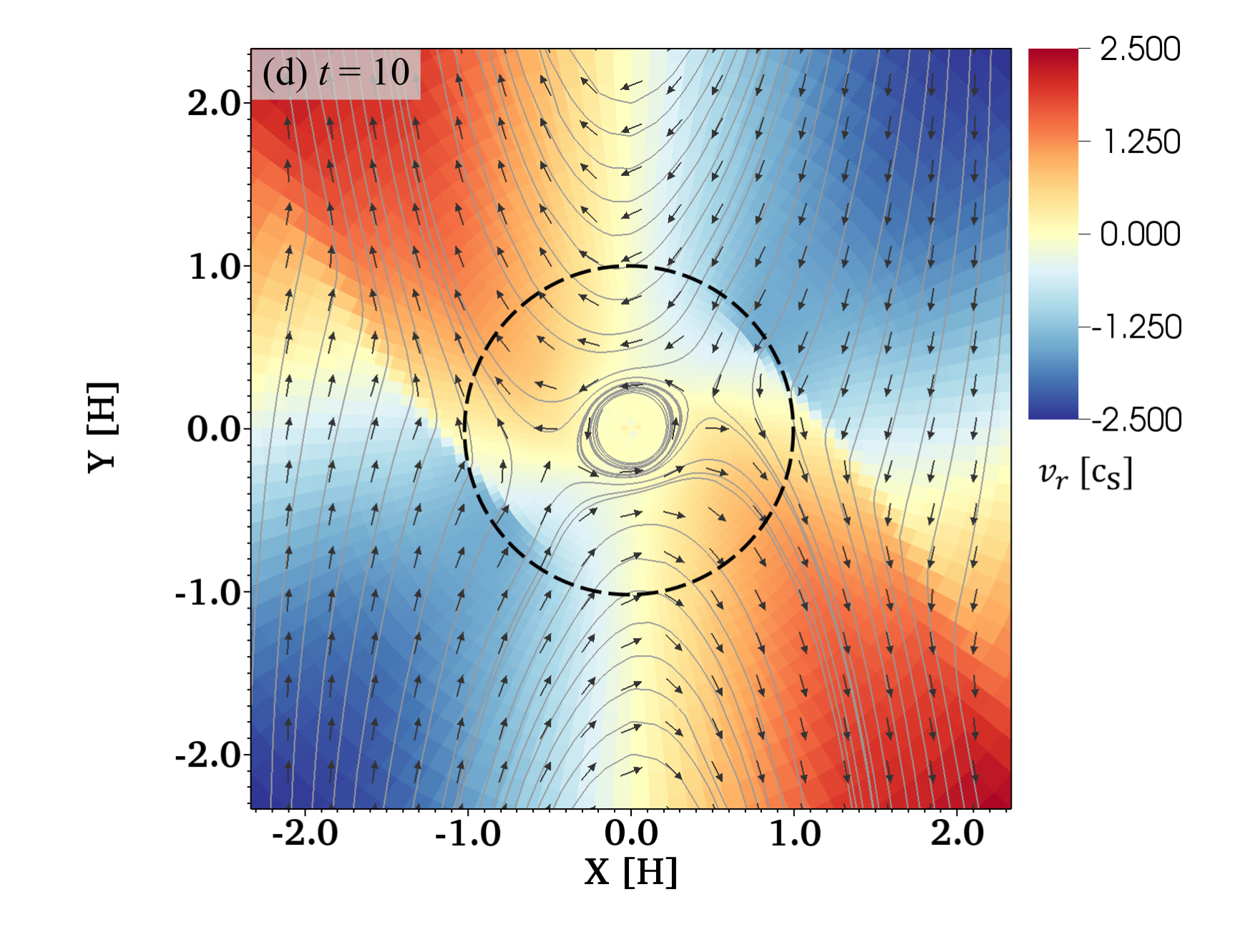}
      \end{minipage} \\
      
 
      \begin{minipage}{0.50\hsize}
        \centering
          \includegraphics[keepaspectratio, width=\linewidth, angle=0]
                          {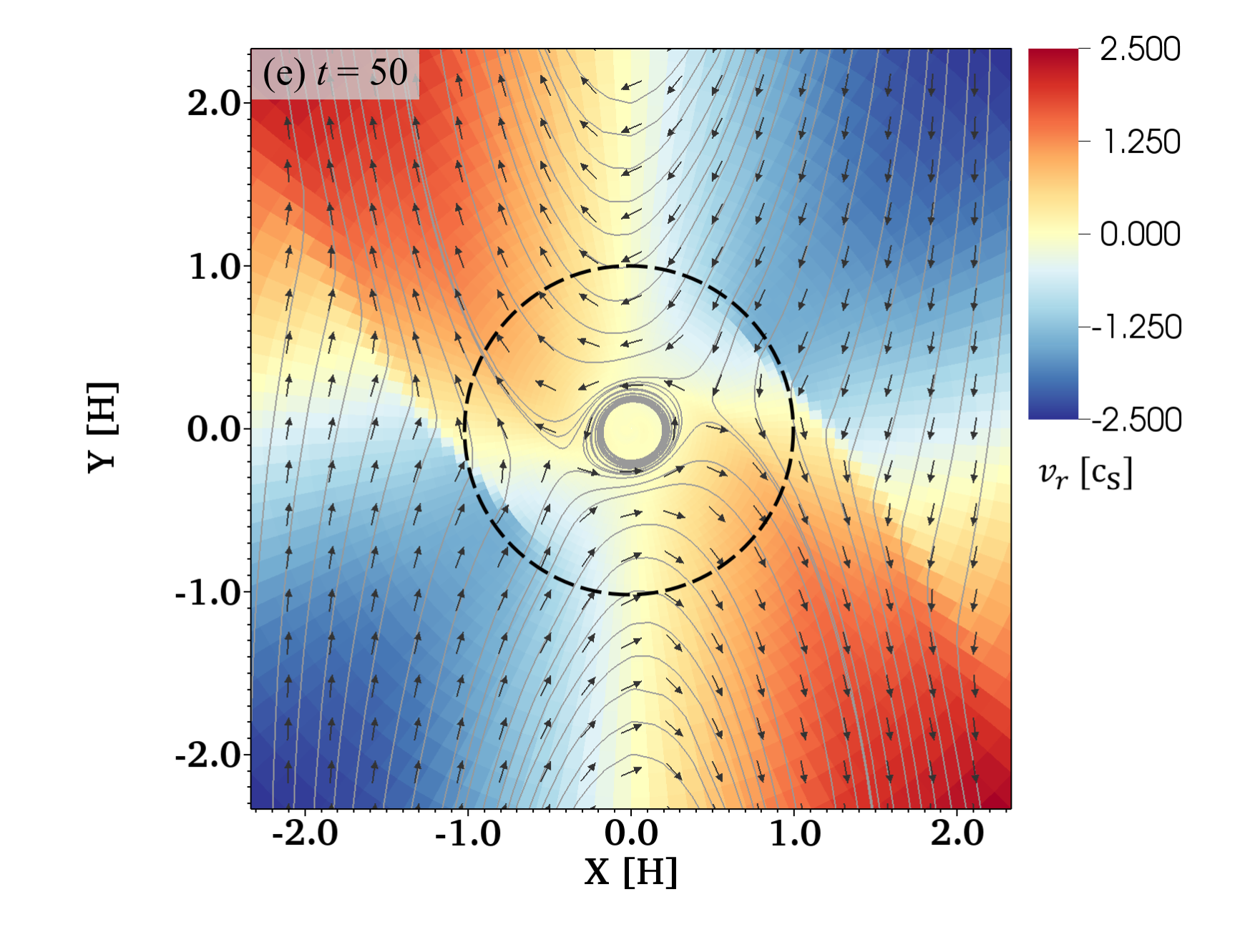}
      \end{minipage}
 

      \begin{minipage}{0.50\hsize}
        \centering
          \includegraphics[keepaspectratio, width=\linewidth, angle=0]
                          {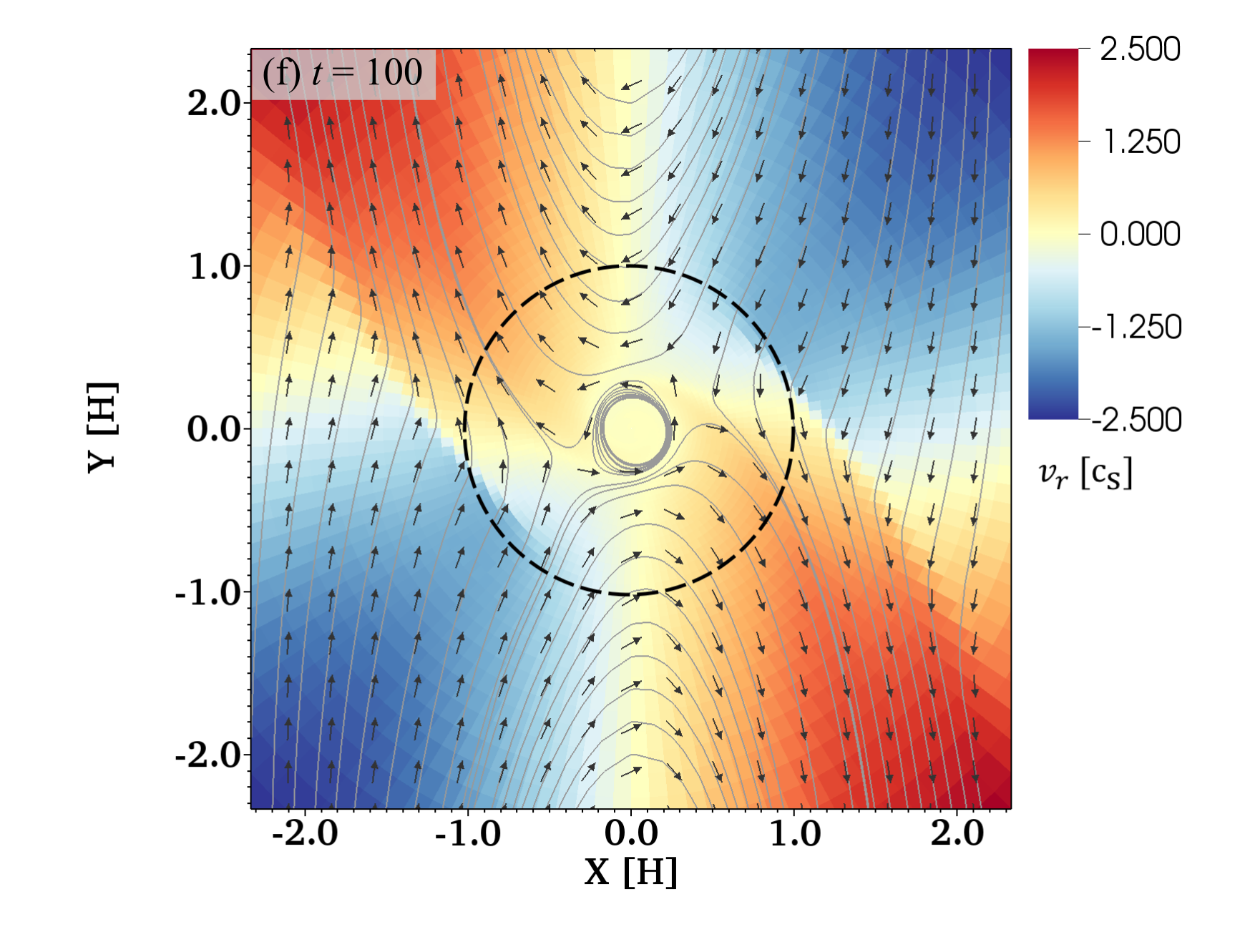}
      \end{minipage} \\
    \end{tabular}
    	\caption{The time evolution of the flow field around a planet having $m=1.0$ at the mid-plane of the disc. Each panel corresponds to the snapshots at $t=2,\ 3,\ 5,\ 10,\ 50,\ \text{and}\ 100$. Since we set the injection time as the unity, in this run, the planet's gravity has completely been inserted into the disc in all panels. Colour contour represents the flow speed in the radial direction. The vertical and the horizontal axis are normalised by the scale height of the disc. The solid and dashed lines correspond to the specific streamlines and the Bondi radius of the planet, respectively. Blue and red regions imply where the gas flows inwards and outwards. After accretion phase ($t<2$), gas flowing from the centre of the planet (outflow) emerges in the early stage of the time evolution. The topologies of the flow field have not significantly changed after $t=5$ in this simulation. We note that the length of the arrows does not scale with the flow speed.}
	\label{fig:mid-plane}
\end{figure*}

\begin{figure*}[htpb]
  \centering
    \begin{tabular}{c}
 
 
      \begin{minipage}{0.50\hsize}
        \centering
          \includegraphics[keepaspectratio, width=\linewidth, angle=0]
                          {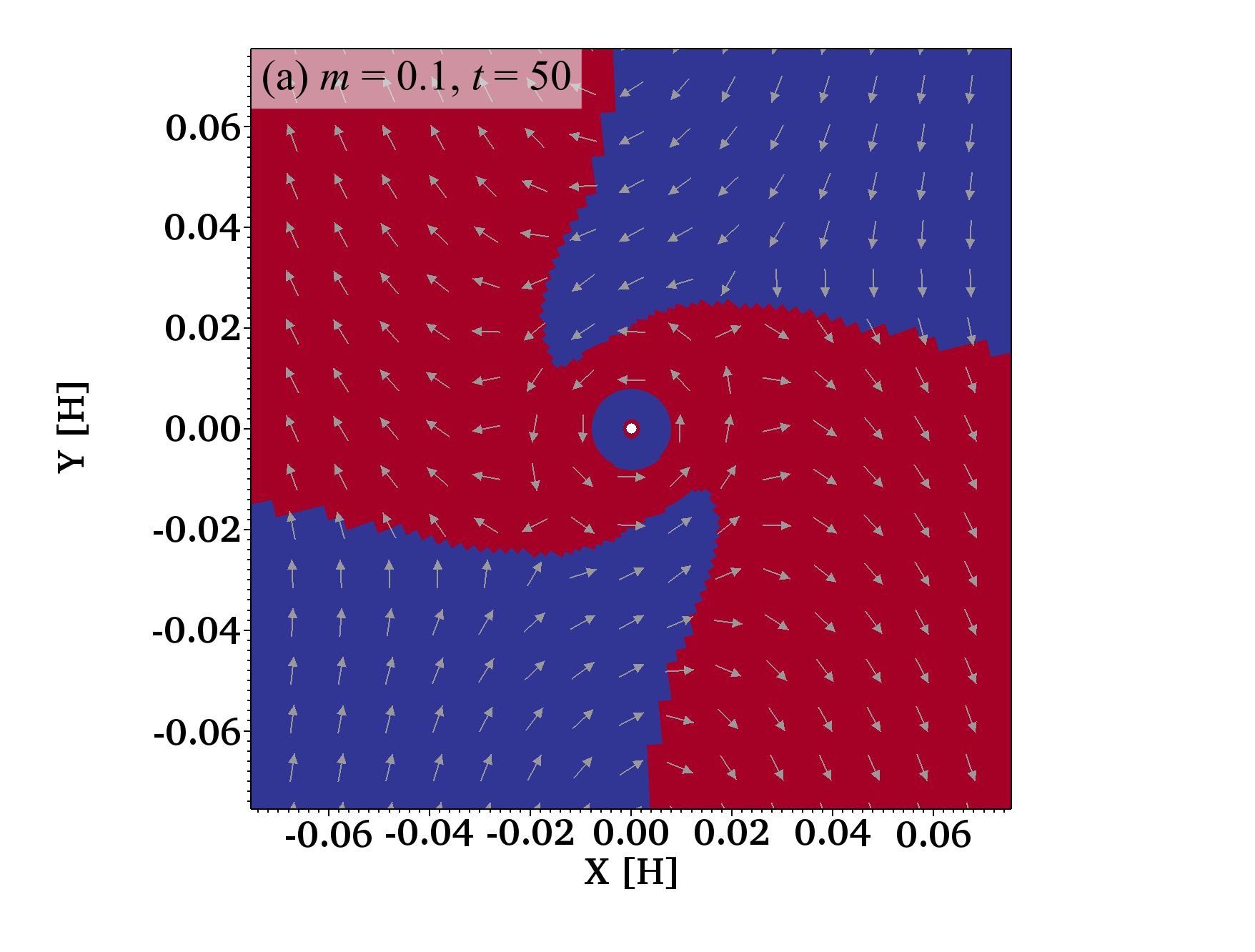}
      \end{minipage}
 
 
      \begin{minipage}{0.50\hsize}
        \centering
          \includegraphics[keepaspectratio, width=\linewidth, angle=0]
                          {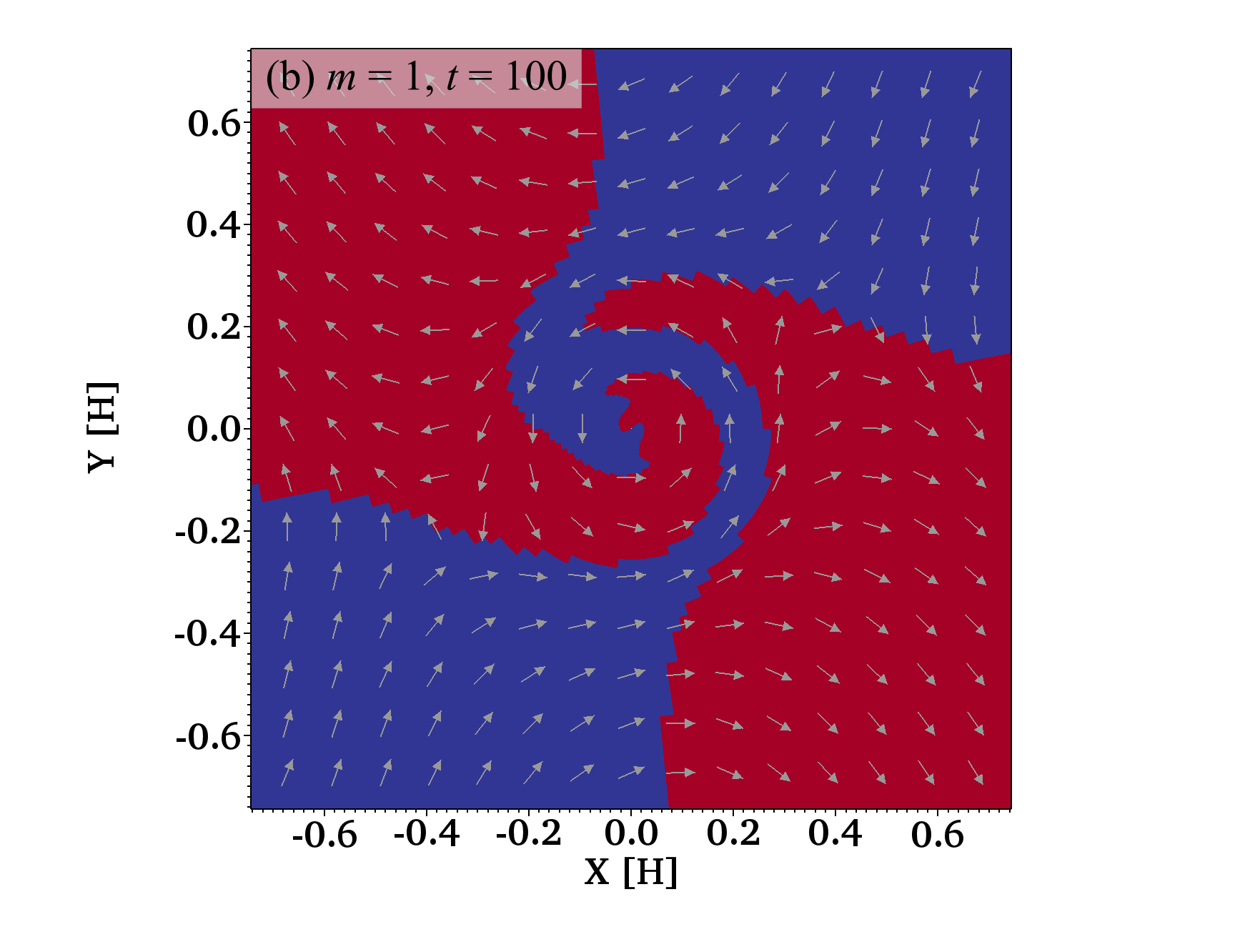}
      \end{minipage} \\
    \end{tabular}
\caption{The structure of the inflow and outflow in the region close to the planet at the mid-plane. (a): The result obtained from \texttt{m01} at $t=50$. (b): The result obtained from \texttt{m1} at $t=100$. In order to distinguish the inflow and outflow clearly, we plotted $v_{r}/|v_{r}|$. Red and blue correspond to the regions where $v_{r}/|v_{r}|$ has positive and negative values. We note that the length of the arrows does not scale with the flow speed.}
\label{fig:vicinity}
\end{figure*}   

 \begin{figure}[htbp]
 \resizebox{\hsize}{!}
 {\includegraphics[width=\linewidth]{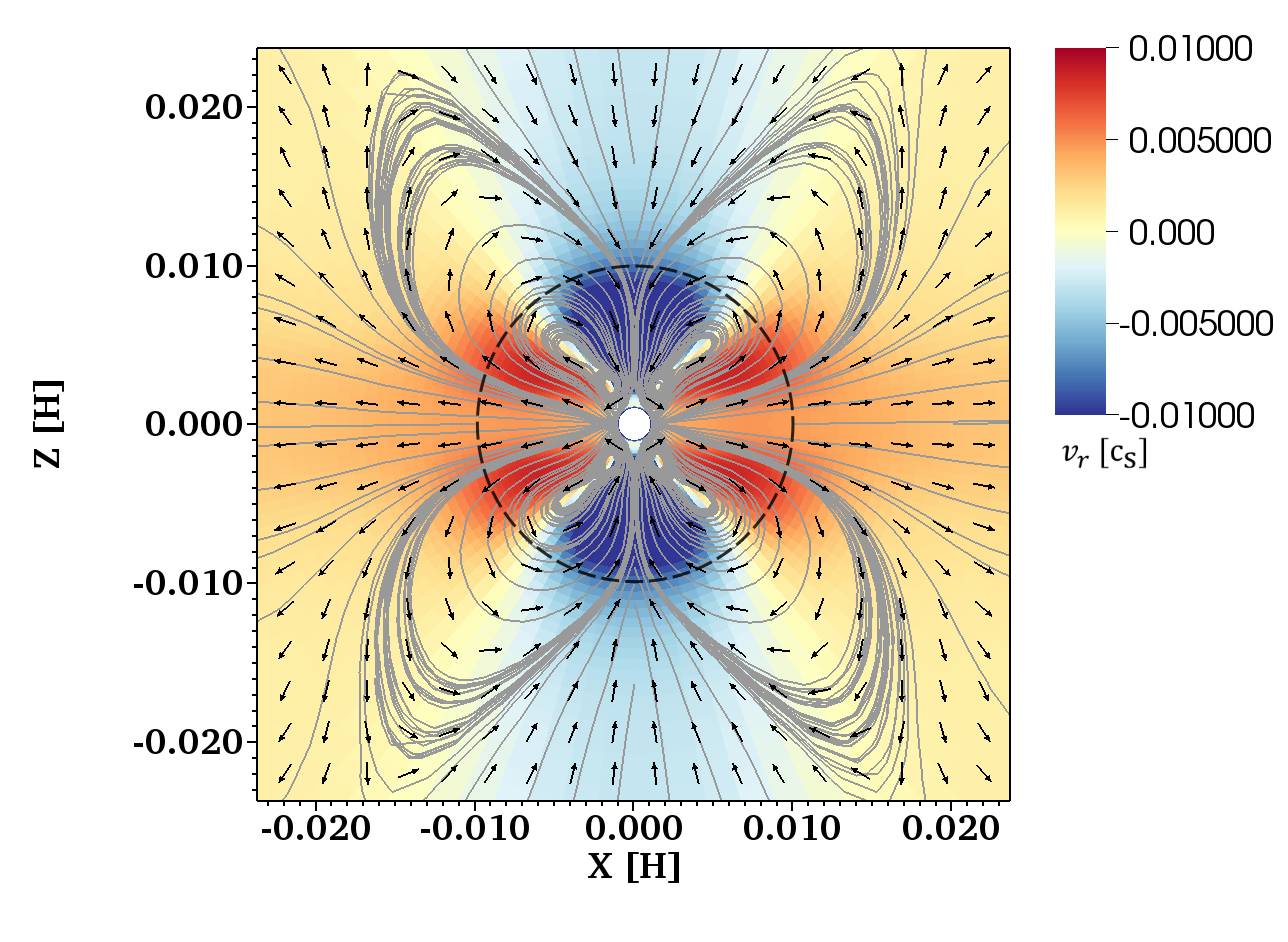}} 
 \caption{The vertical structure of inflow and outflow at the meridian plane ($y=0$) obtained from \texttt{m001} at $t=10$. The red or blue region means where gas flows outwards or inwards in the radial direction. The solid and the dashed lines represent the specific streamlines of gas flow and the Bondi radius of the planet, respectively. The meaning of the contour is the same as that in \Figref{fig:mid-plane}, but the ranges are different. Since we adjusted the range of contour to show the vertical structure of the outflow, the value of inflow is saturated in the vicinity of the planet. We note that the length of the arrows does not scale with the flow speed.}
\label{fig:meridian01}
\end{figure}

\begin{figure*}[!htbp]
  \centering
    \begin{tabular}{c}
 
 
      \begin{minipage}{0.50\hsize}
        \centering
          \includegraphics[keepaspectratio, width=\linewidth, angle=0]
                          {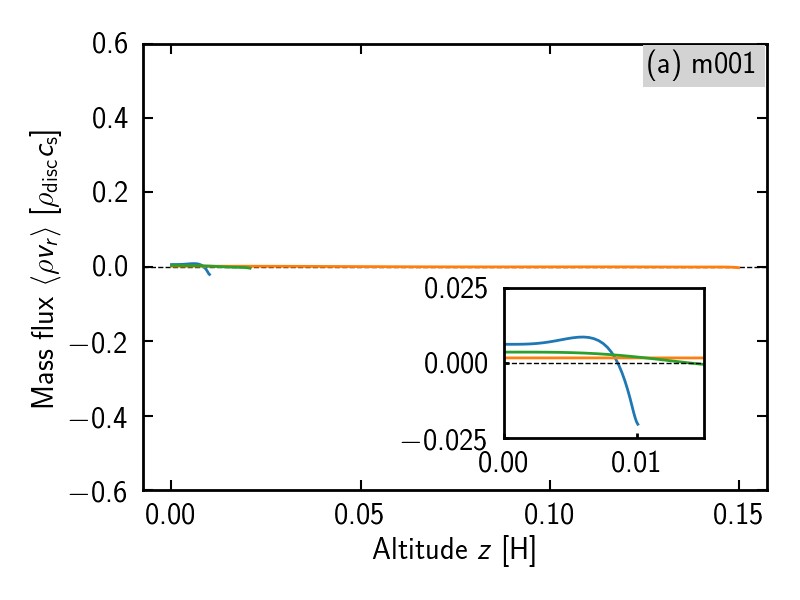}
      \end{minipage}
 
 
      \begin{minipage}{0.50\hsize}
        \centering
          \includegraphics[keepaspectratio, width=\linewidth, angle=0]
                          {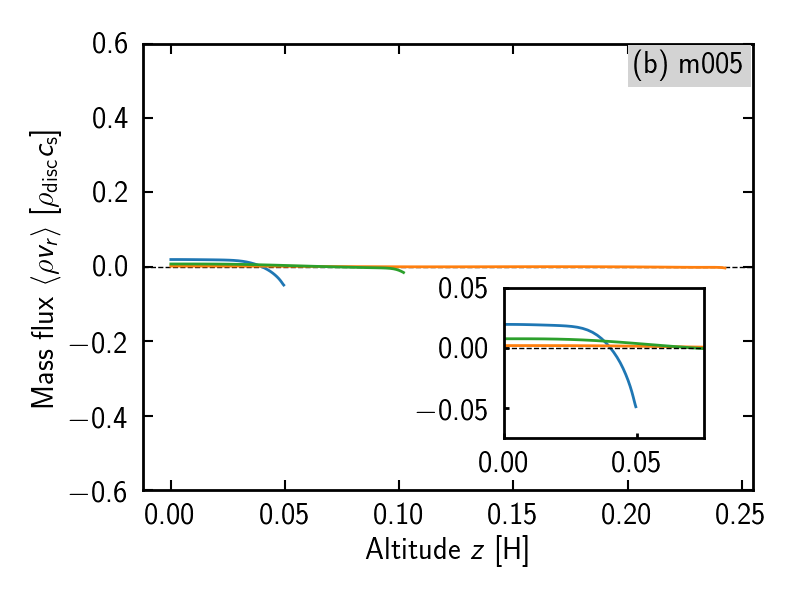}
      \end{minipage} \\
       
 
      \begin{minipage}{0.50\hsize}
        \centering
          \includegraphics[keepaspectratio, width=\linewidth, angle=0]
                          {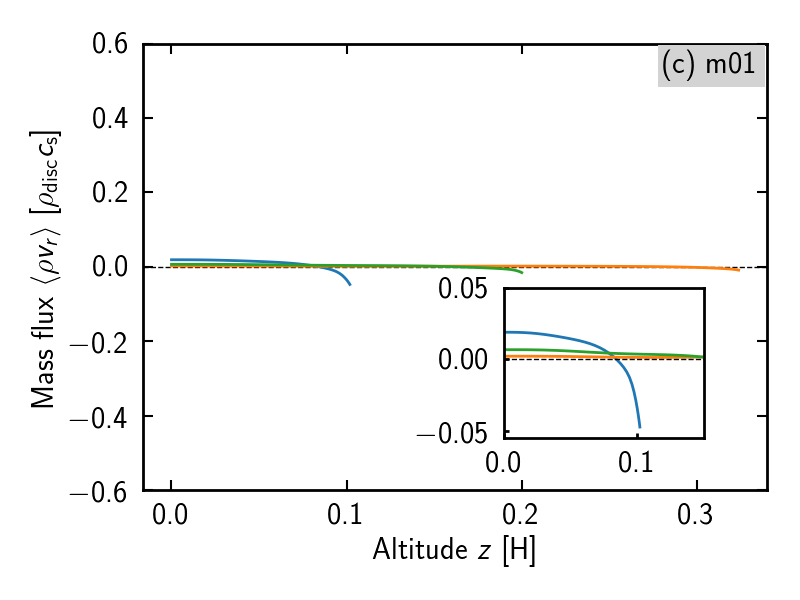}
      \end{minipage}
 
 
      \begin{minipage}{0.50\hsize}
        \centering
          \includegraphics[keepaspectratio, width=\linewidth, angle=0]
                          {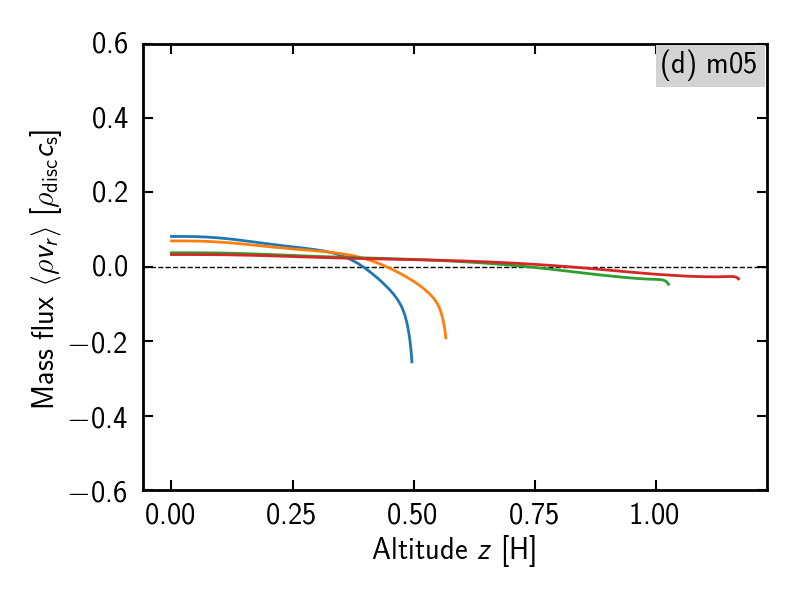}
      \end{minipage} \\
       
 
      \begin{minipage}{0.50\hsize}
        \centering
          \includegraphics[keepaspectratio, width=\linewidth, angle=0]
                          {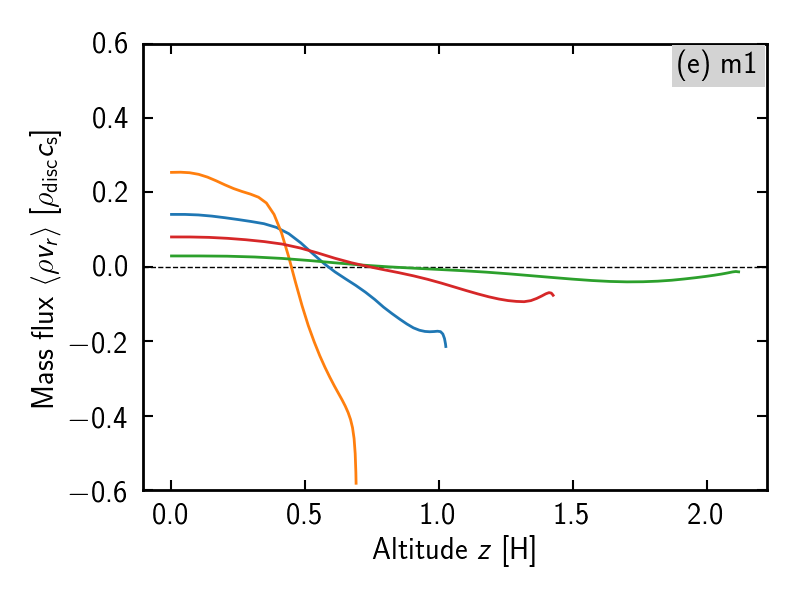}
      \end{minipage}
 
 
      \begin{minipage}{0.50\hsize}
        \centering
          \includegraphics[keepaspectratio, width=\linewidth, angle=0]
                          {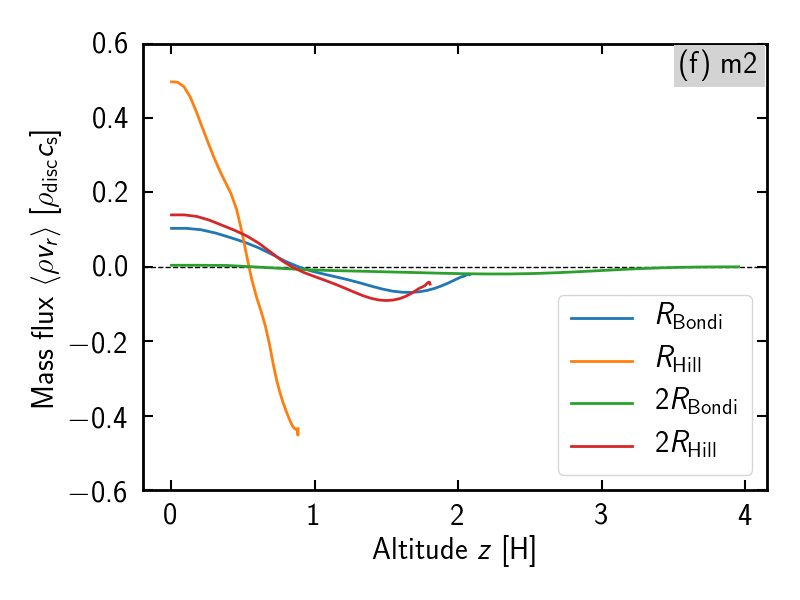}
      \end{minipage}
    \end{tabular}
\caption{The azimuthally averaged mass flux, $\langle\rho v_{r}\rangle_{\phi}$, as a function of altitude, $z$. Each panel shows each result obtained from analysis of a series of simulations labelled in the upper right. Each solid line represents the changes of $\langle\rho v_{r}\rangle_{\phi}$; blue: altitude is varied along with the Bondi radius of the planet; orange: along with the Hill radius; green: along with the twice of the size of the Bondi radius; red: along with the twice of the size of the Hill radius. The correspondences of each solid line are shown in the lower right legend of (f). The figures displayed at the lower right of (a)-(c) are an enlarged view of a specific area in each panel. Gas flows in where $\langle\rho v_{r}\rangle_{\phi}<0$ and flows out where $\langle\rho v_{r}\rangle_{\phi}>0$, respectively. Black dashed line corresponds to $\langle\rho v_{r}\rangle_{\phi}=0$ where inflow and outflow are balanced. We assume substantial inflow and outflow occur in the area where mass flux is dominant; that is, gas mainly flows in and out of the Bondi or Hill sphere of the planet, depending on which is smaller. We note that the scale of the vertical axis is the same, but the horizontal axis is different for each panel.}
\label{fig:massflux}
\end{figure*}   

\begin{figure*}[!htbp]
  \centering
    \begin{tabular}{c}
 
 
      \begin{minipage}{0.50\hsize}
        \centering
          \includegraphics[keepaspectratio, width=\linewidth, angle=0]
                          {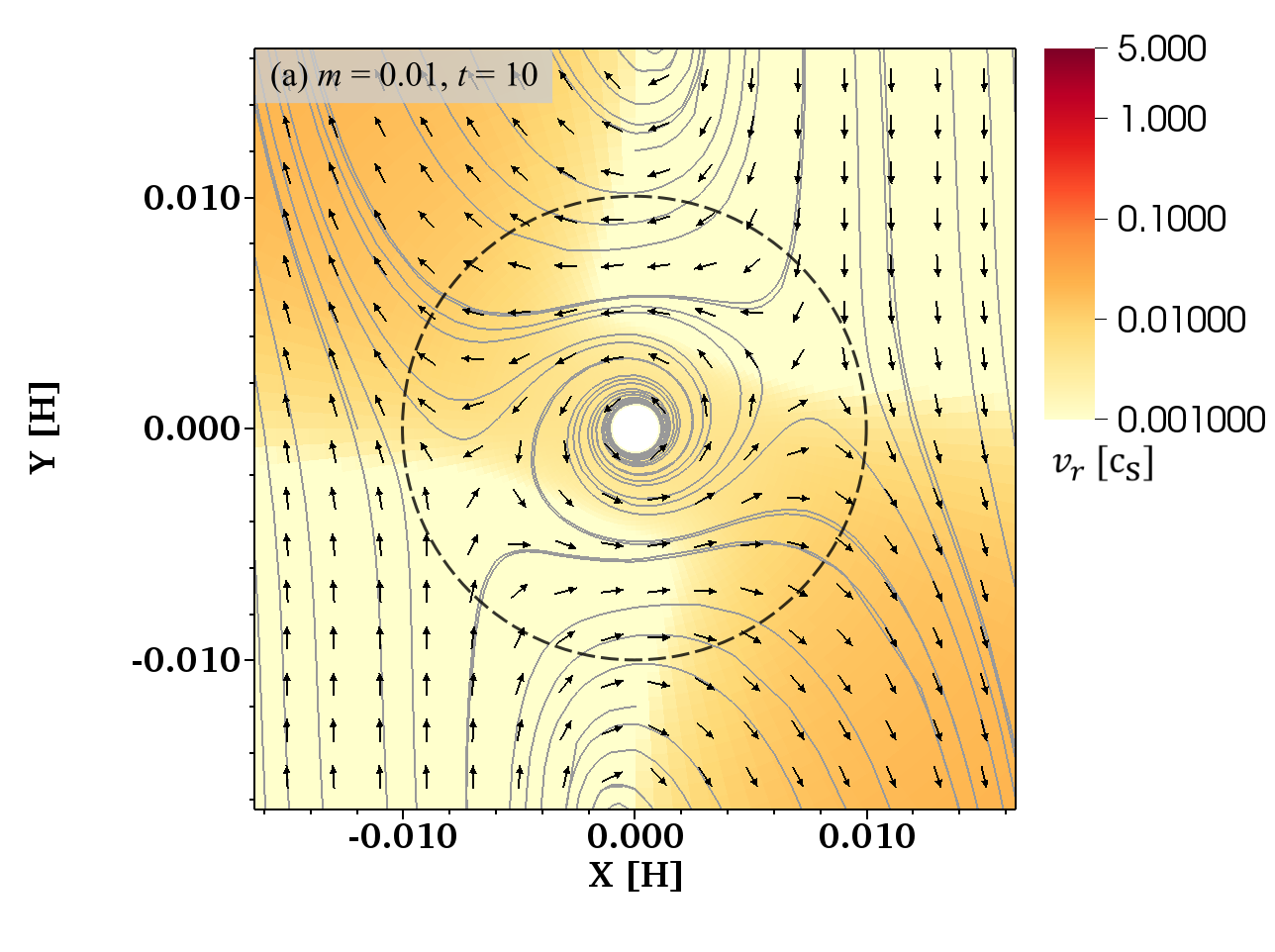}
      \end{minipage}
 
 
      \begin{minipage}{0.50\hsize}
        \centering
          \includegraphics[keepaspectratio, width=\linewidth, angle=0]
                          {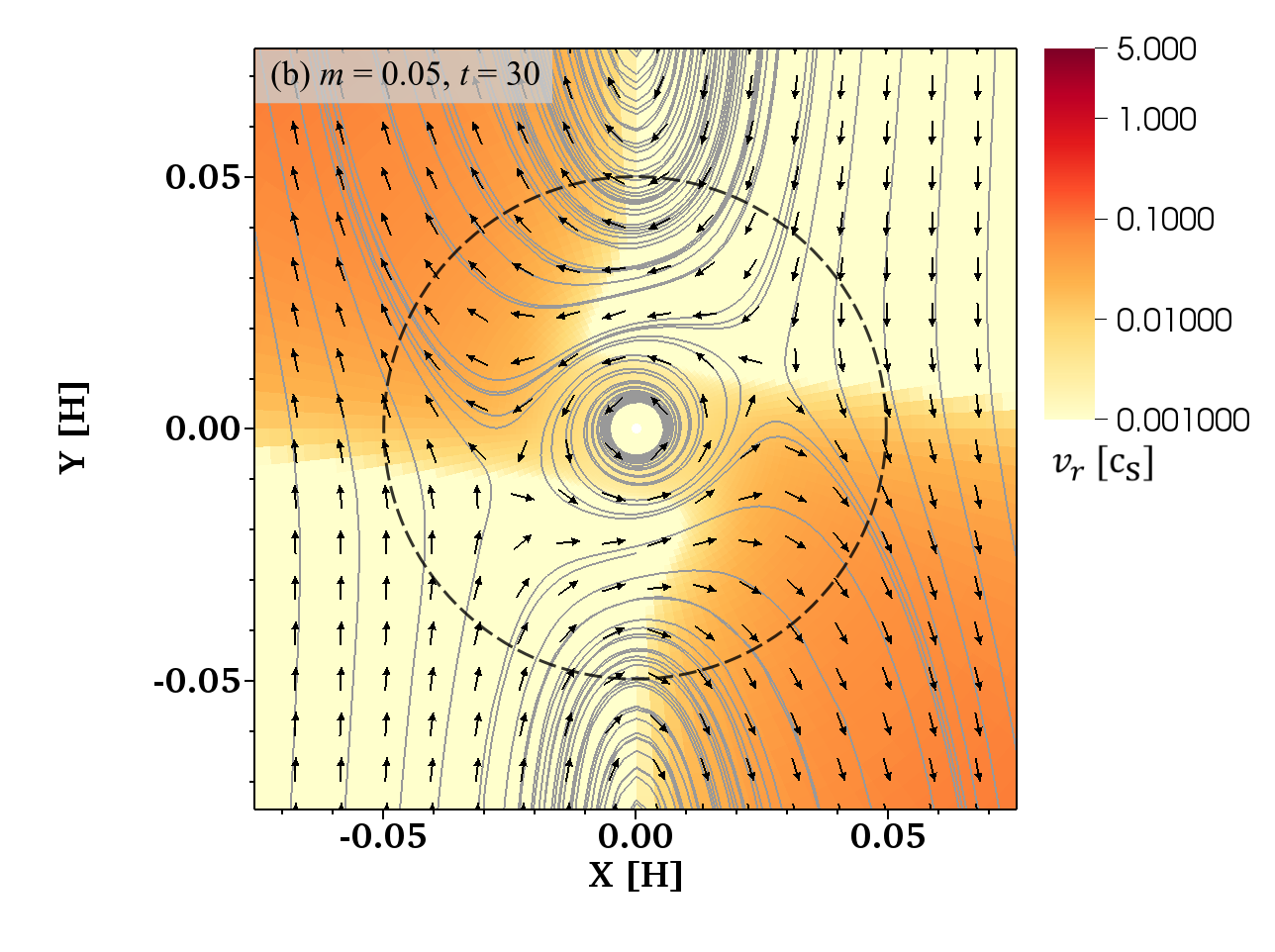}
      \end{minipage} \\
 
 
      \begin{minipage}{0.50\hsize}
        \centering
          \includegraphics[keepaspectratio, width=\linewidth, angle=0]
                          {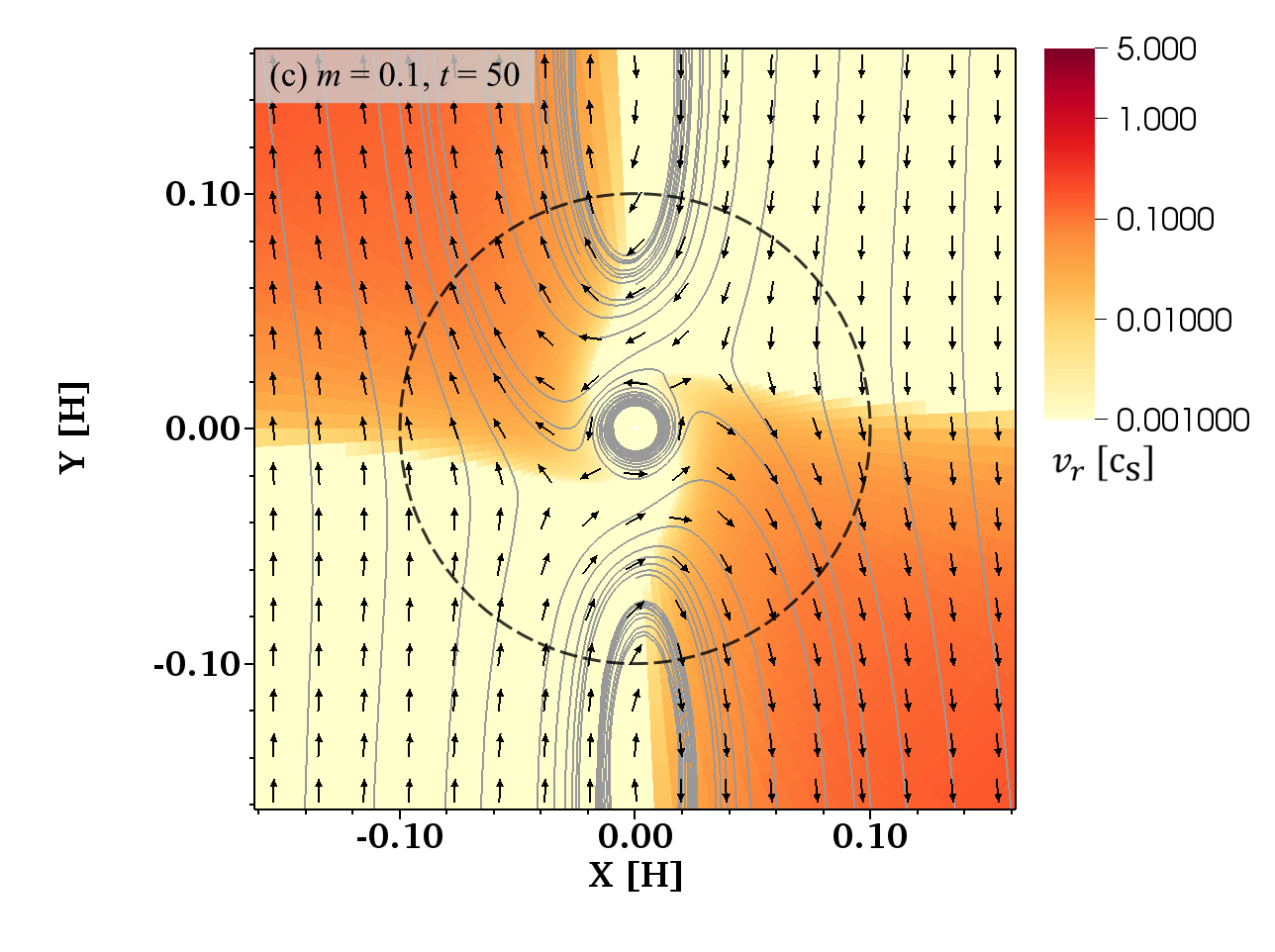}
      \end{minipage}
 
 
      \begin{minipage}{0.50\hsize}
        \centering
          \includegraphics[keepaspectratio, width=\linewidth, angle=0]
                          {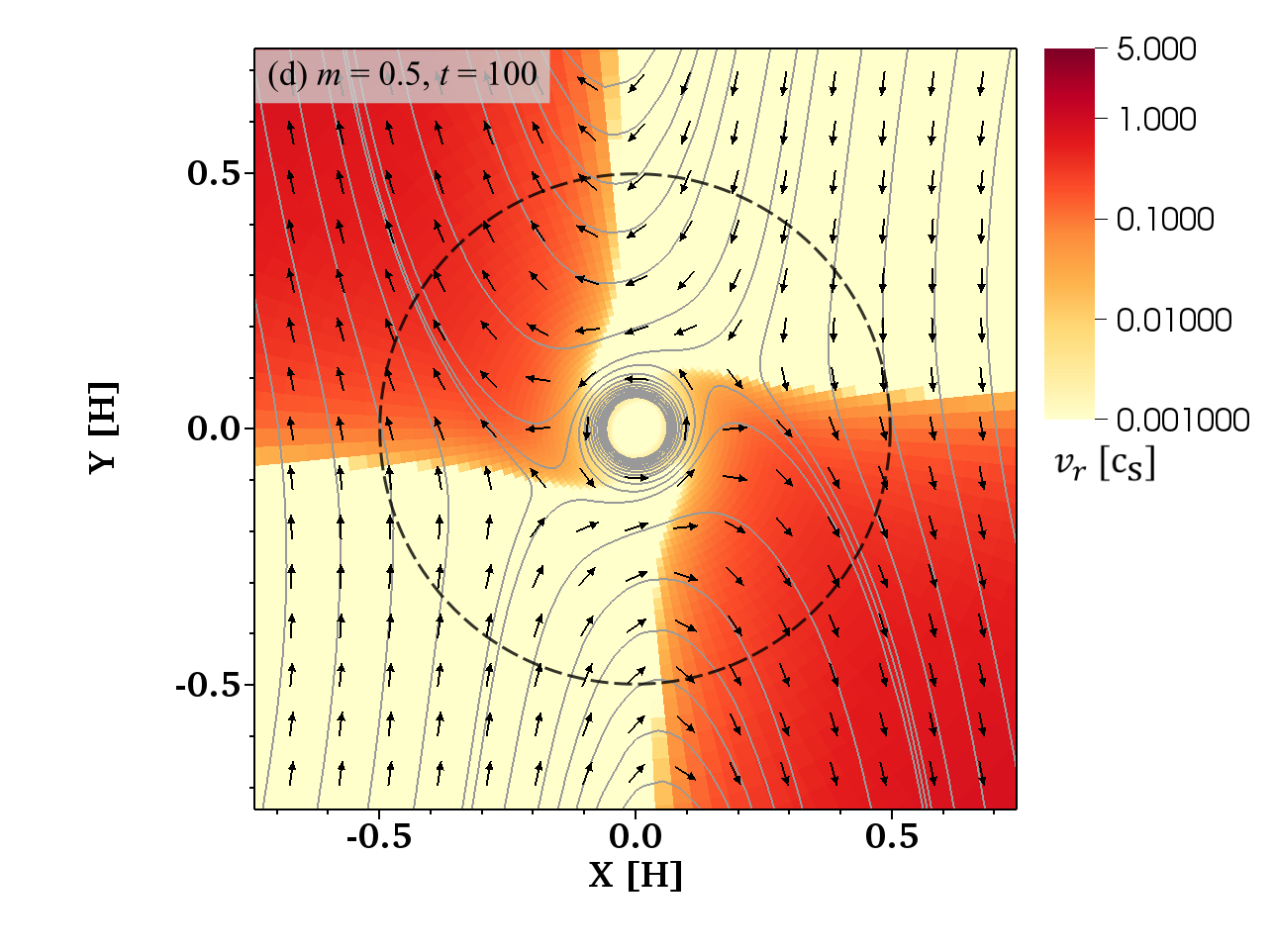}
      \end{minipage} \\
      
 
      \begin{minipage}{0.50\hsize}
        \centering
          \includegraphics[keepaspectratio, width=\linewidth, angle=0]
                          {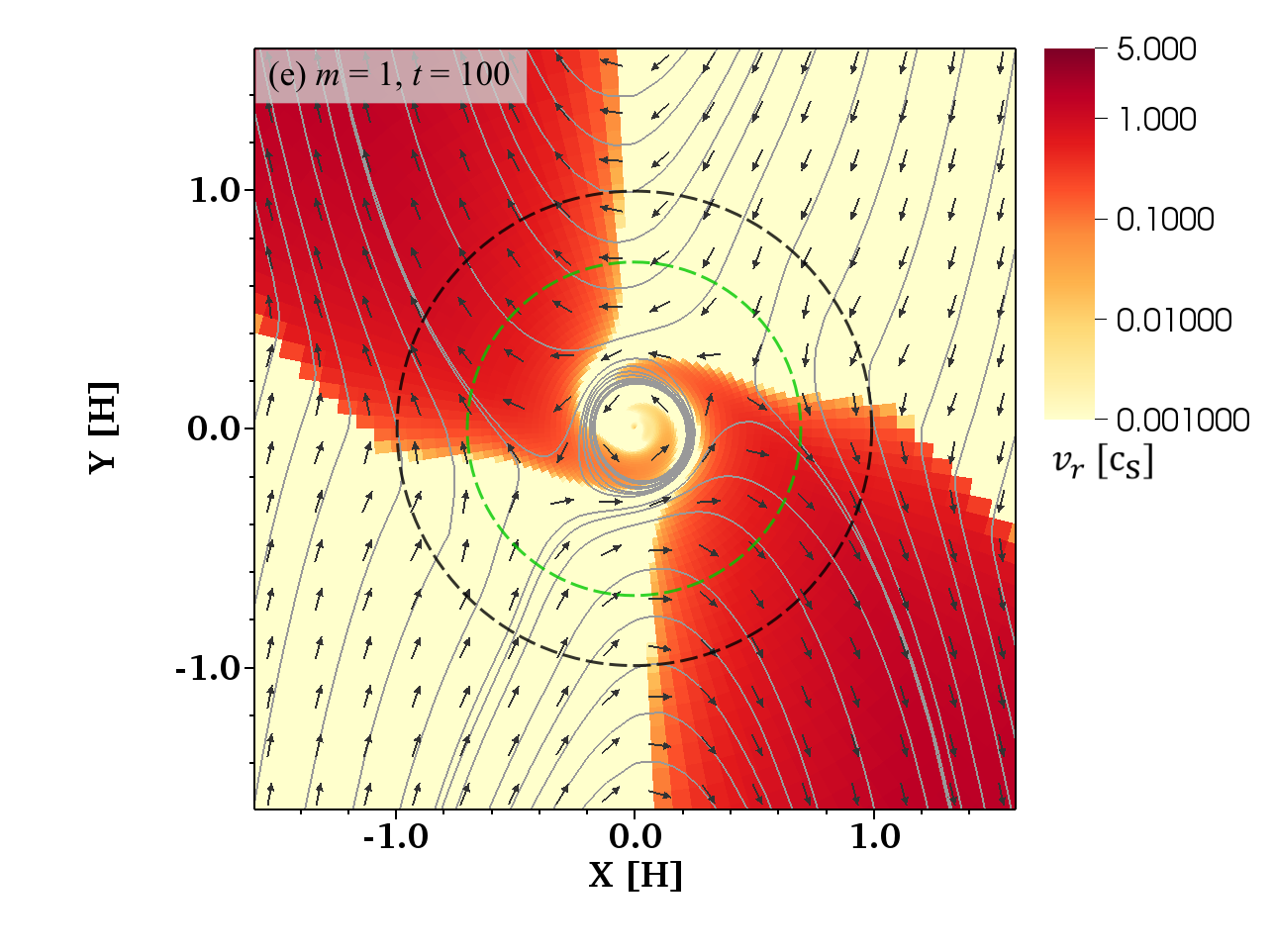}
      \end{minipage}
 

      \begin{minipage}{0.50\hsize}
        \centering
          \includegraphics[keepaspectratio, width=\linewidth, angle=0]
                          {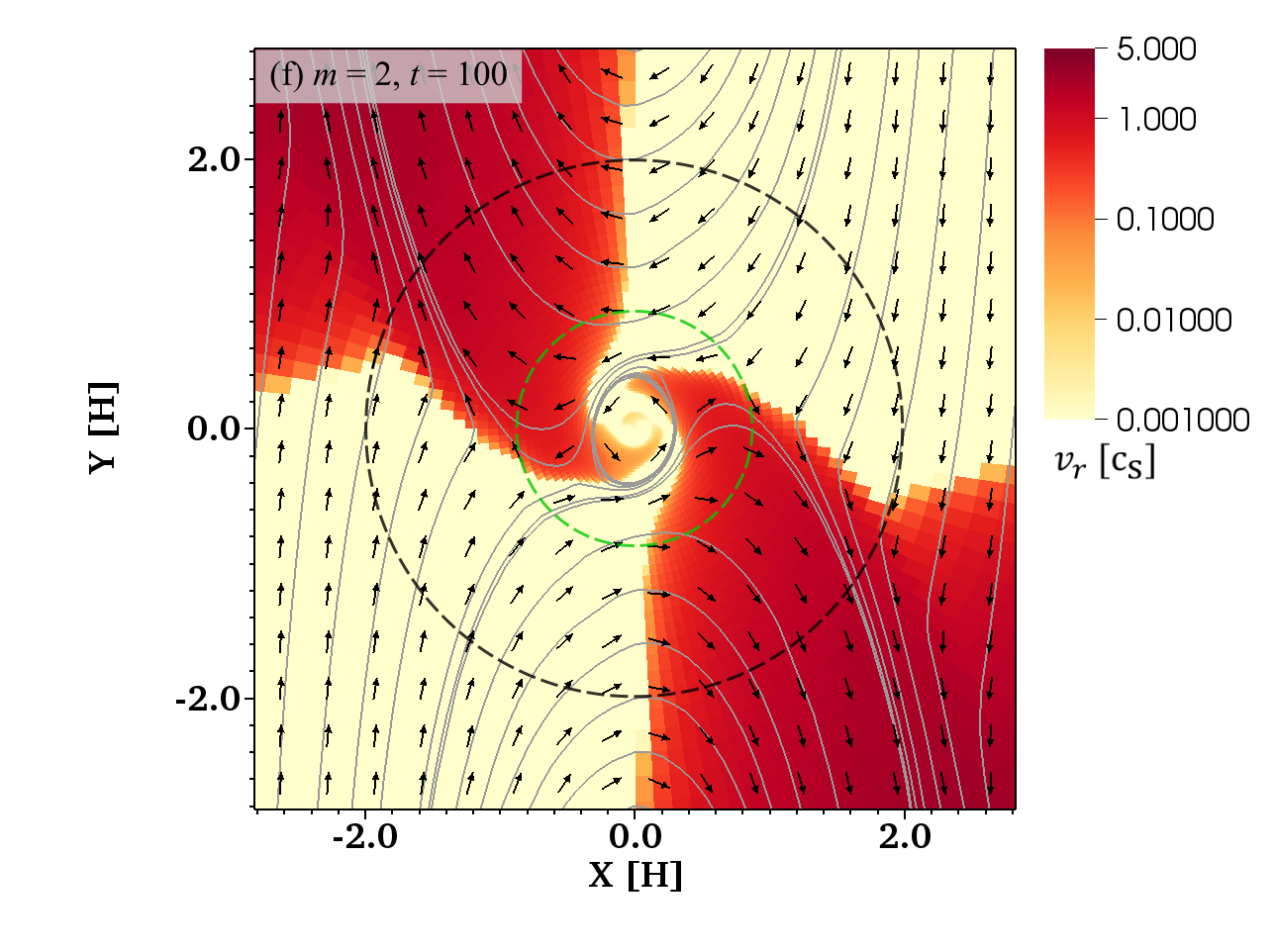}
      \end{minipage} \\
    \end{tabular}
    	\caption{The changes of the outflow speed at the mid-plane of the disc according to the planetary mass. Each panel corresponds to each result obtained from different simulations at $t=t_{\rm end}$. Colour contour represents the flow speed in the radial direction. In order to show the differences of the outflow speed specifically, we plotted in a logarithmic scale. We note that the colour contour is saturated for the inflow ($v_{r}<0$) region. The orange-red or yellow areas mostly correspond to the region where the radial velocity is positive or negative (see \Figref{fig:mid-plane}). The solid lines correspond to the specific streamlines of the gas flow. The black dashed line corresponds to the Bondi radius of the planet. The green dashed line shown in panel (e) and (f) corresponds to the Hill radius. We note that the length of the arrows does not scale with the flow speed.}
	\label{fig:outflowlog}
\end{figure*}
\begin{figure*}[!htbp]
  \centering
    \begin{tabular}{c}
 
 
      \begin{minipage}{0.50\hsize}
        \centering
          \includegraphics[keepaspectratio, width=\linewidth, angle=0]
                          {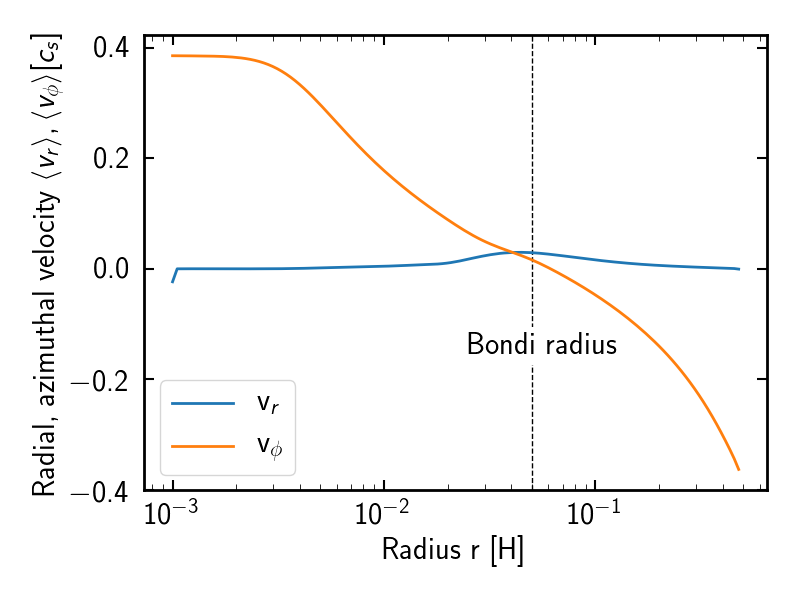}
      \end{minipage}
 
 
      \begin{minipage}{0.50\hsize}
        \centering
          \includegraphics[keepaspectratio, width=\linewidth, angle=0]
                          {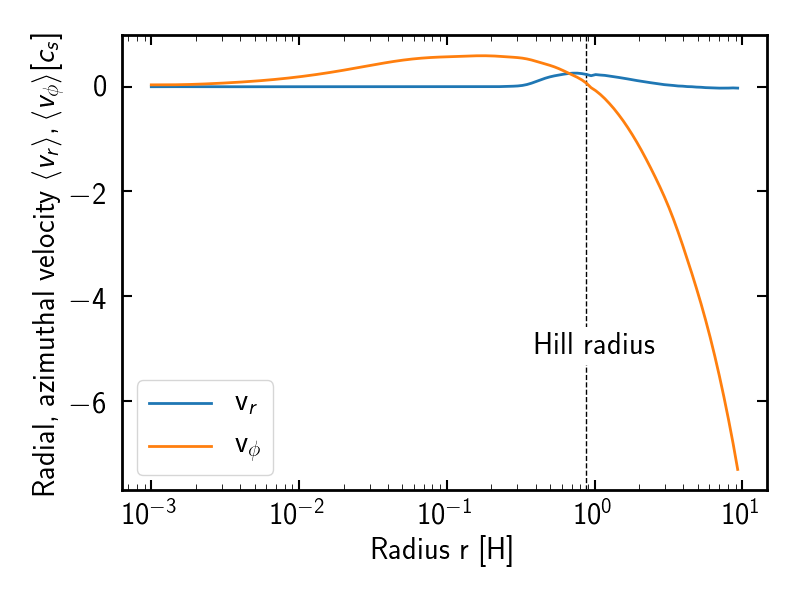}
      \end{minipage} 
    \end{tabular}
\caption{The azimuthally averaged radial and azimuthal velocity as a function of the radius $r$. The left panel: the result from \texttt{m005} run at $t=30$.  The right panel: the result from \texttt{m2} run at $t=100$. The solid lines correspond to the radial (blue) and azimuthal (orange) velocity of the gas flow in the vicinity of the planet. The dashed line shows the position of the Bondi or Hill radius of the planet.}
\label{fig:velrp}
\end{figure*}   

 \begin{figure}[htbp]
 \resizebox{\hsize}{!}
 {\includegraphics{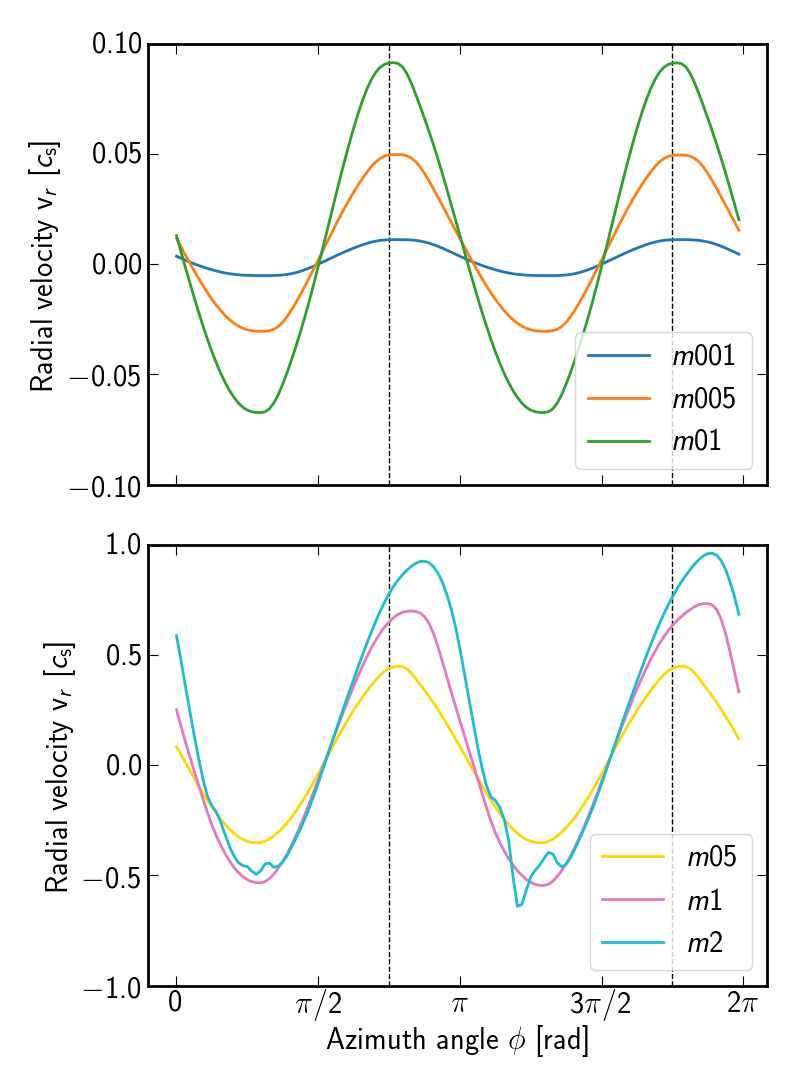}} 
 \caption{The changes of the radial velocity in the azimuth direction at the Bondi (for \texttt{m001, m005, m01, m05}) or Hill radius (for \texttt{m1, m2}) on the mid-plane of the disc at $t=t_{\rm end}$. The solid lines correspond to each analytical result obtained from each simulation. Each dashed line corresponds to the locations where $\phi=3\pi/4,\ 7\pi/4$. The radial velocity has the maximum value near these points after it takes the periodic oscillation. The maximum value of $v_{r}$ increases with the planetary mass. The positive or negative value of $v_{r}$ means where inflow or outflow emerges. }
\label{fig:vrmax}
\end{figure}

 \begin{figure}[htbp]
 \resizebox{\hsize}{!}
 {\includegraphics{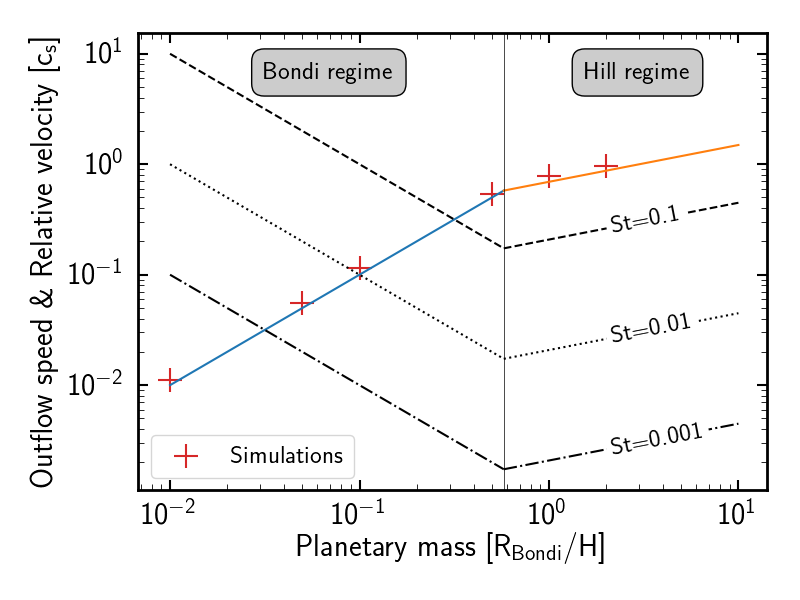}} 
 \caption{The dependency of the outflow speed as a function of the dimensionless planetary mass. The cross symbols are the maximum value of the outflow speed at the Bondi (for \texttt{m001, m005, m01, m05}) or Hill radius (for \texttt{m1, m2}) on the mid-plane of the disc obtained from our simulations at $t=t_{\rm end}$. The blue  and orange solid lines correspond to the analytical solution of the outflow speed derived from \cref{eq:bondiregime,eq:hillregime}, which are divided at the point where the size of the Bondi radius of the planet exceeds the size of the Hill radius (see subsection \ref{sec:outflow} for details). Each black line shows the changes of the relative velocity of the small particles for each Stokes number, ${\rm St}=10^{-3}$ (dashed-dotted line), $10^{-2}$ (dotted line), and $10^{-1}$ (dashed line) derived from \cref{eq:relativebondi,eq:relativehill} (see subsection \ref{sec:relativevelocity} for details). }
\label{fig:outflow}
\end{figure}

\begin{figure*}[!htbp]
\centering
\includegraphics[width=\linewidth]{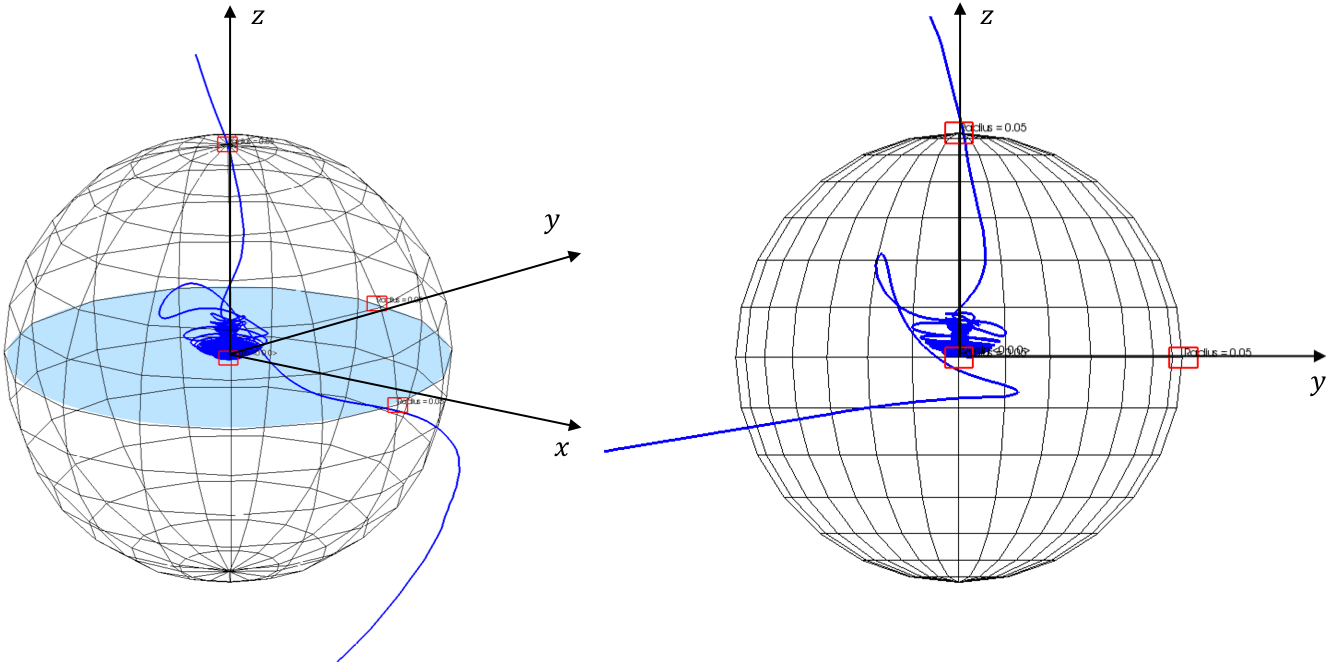}
\caption{A single specific streamline of gas flow from \texttt{m005} run at $t=10$. The sphere in this figure represents the Bondi sphere of the planet. The blue solid line indicates the streamline which enters at the zenith of the Bondi sphere of the planet and circles around the planet several times, finally exiting near the mid-plane of the disc, which is coloured light blue. For convenience when plotting a streamline, inflow point is located at $(x,y,z)=(0.001, -0.001, 0.05)$, which is slightly different from $P_{1}=(0,0,0.05)$ (see subsection \ref{sec:outflow}).}
\label{fig:3Dsinglestreamline}
\end{figure*}

\section{Results} \label{sec:result}
The main subject of this study is to clarify the dependence of the gas flow field on the planetary mass. As a result of a series of simulations, there were some properties that varied depending on the mass of the planet and others that did not. In subsection \ref{sec:result1}, we show universal properties of the flow field independent of the planetary mass. Subsection \ref{sec:result2} shows the dependence of the flow field on the planetary mass.

\subsection{The structure of the gas flow field}\label{sec:result1}
Through performing three-dimensional hydrodynamical simulations, we found the characteristic structures of the flow field around the planets. Figure \ref{fig:3Da} shows the three-dimensional structure of the flow field around the embedded planet obtained from calculation of the \texttt{m1} run at $t=100$. Figure \ref{fig:3Da}a is a bird's-eye view of the flow field. Figure \ref{fig:3Da}b and c are the $x$-$y$ plane viewed from $+z$ direction and the $y$-$z$ plane viewed from $-x$ direction, respectively. A planet is located at the centre of the Bondi sphere which is expressed by the black solid lines in each panel. The structure of the flow is consistent with previous studies \cite[]{Ormel:2015b,Fung:2015,Cimerman:2017,Lambrechts:2017,Kurokawa:2018}. Gas flows in at high latitudes of the Bondi sphere of the planets and leaves through the mid-plane of the disc. The flow field has three types of streamlines defined as follows.

\textit{The Keplerian shear streamline}: The Keplerian shear exists near $x\gtrsim1.5$ and extends in the $y$-direction (\Figref{fig:3Da}a,b). These streamlines are slightly perturbed by the planet's gravity, however, and do not accrete onto the planet.

\textit{The horseshoe streamline}: The horseshoe flow exists in the anterior-posterior direction of the planet’s orbital direction (\Figref{fig:3Da}a,b). The streamlines show a columnar structure as reported in \cite{Fung:2015}. In all of our simulations, we found the horseshoe flow had a similar vertical structure. The streamlines that are relatively close to the planet underwent a gravitational perturbation by the planet and descended somewhat towards it, but they escaped from the Bondi sphere of the planet without reaching it. 

\textit{The atmospheric recycling streamline}: A few streamlines reach the vicinity of the planet (\Figref{fig:3Da}a-c). These streamlines start to descend halfway along the horseshoe orbit. After entering directly above the planet, however, they are going to sharply descend towards it. They circle the planet several times. One of them ultimately exits the Bondi sphere through the mid-plane region of the disc. The altitude where the gas starts descending is approximately 2 times higher than the top of the Bondi sphere. This streamline connects inside and outside the Bondi sphere. Therefore, this embedded atmosphere represents an open system where gas continuously enters the Bondi sphere and leaves it \cite[]{Ormel:2015b}.

The outflow emerged from the early stage of the time evolution. Figure \ref{fig:mid-plane} shows the time evolution of the flow field at the mid-plane of the disc. A planet is located at the centre of this figure. At the early stage of the time evolution of the flow field, gas accretes onto the planetary core (as shown in \Figref{fig:mid-plane}a). After the accretion phase, the gas began to circulate in the vicinity of the core of the planet. As shown in \Figref{fig:mid-plane}a, inflow (represented in blue) is dominant inside the Bondi sphere at $t<3$.

The situation changes after $t=3$. In Figs. \ref{fig:mid-plane}b-f, there are three types of streamlines: the Keplerian shear streamlines, the horseshoe streamlines, and the atmospheric recycling streamlines. The Keplerian shear exists on both the left and right sides of the planet. Although it is slightly distorted due to the gravity of the planet, its trajectory is almost straight without accreting onto the core of the planet. Horseshoe flows are rotationally symmetric with respect to the $z$-axis. Part of the horseshoe flow enters the Bondi sphere of the planet and exits it, drawing a U-turn curve, without accreting onto the core of the planet. Atmospheric recycling streamlines connect inside and outside the Bondi sphere of the planet. In this phase, the topology of the flow field does not change significantly with time in the \texttt{m1} run. The shape of the streamlines experiences only a slight change in the vicinity of the planet. A similar trend was also confirmed regardless of the assumed planetary mass. After the accretion phase, the flow field approached the steady state. 

What is striking about \Figref{fig:mid-plane} is that the gas flows out from the core of the planet towards an area outside the Bondi region, travelling along the horseshoe streamlines or atmospheric recycling streamlines at the mid-plane of the disc. Other simulations assuming different planetary masses show similar tendencies of the outflow of gas from the vicinity of the planet. In \Figref{fig:mid-plane}, the speed of outflow near the Bondi radius has reached the order of the isothermal sound speed. Such a fast flow of gas at the mid-plane of the disc has the potential to affect the solid materials around the planet. If the outflow speed is high compared to the radial drift speed of the solid materials, it is probable that outflow behaves as a barrier against the accretion of solid materials and suppresses the accretion rate.

In the region close to the planet, the radial outward flow is dominant (Figs. \ref{fig:vicinity}a and b). We confirmed the dominance of outflow in the mid-plane in the vicinity of the planet in all of our simulations. When $m\leq0.58$ ($R_{\rm Bondi}<R_{\rm Hill}$), inflow emerged only at $r\gtrsim0.1$--$0.2R_{\rm Bondi}$ (\Figref{fig:vicinity}a). Although inflow tails intrude deep inside the Bondi sphere when $m\geq0.58$ ($R_{\rm Bondi}>R_{\rm Hill}$), their width is narrow (\Figref{fig:vicinity}b). The outflow in the mid-plane near the planet is expected to reduce the accretion of solid materials (see subsection \ref{sec:relativevelocity}).

Figure \ref{fig:meridian01} shows the vertical structure of inflow and outflow at the meridian plane, $y=0$. The solid lines are the specific streamlines and they have an axisymmetric structure. Gas flows in from the vertical direction of the $z$-axis and escapes near the mid-plane to mid-latitudes of the disc. In this figure, the gas seems to be descending from a height twice as high as the top of the Bondi sphere or more. Some of the streamlines correspond to the outflow circulating inside the Bondi sphere. The vertical scale of the outflow is about $0.5 R_{\rm Bondi}$. The vertical scale of the outflow is important because the influence on the dust or pebble accretion would be determined by the ratio of the vertical extent of the flow to the scale height of the solid materials.

The fundamental features of the flow field introduced above were not changed significantly with planetary mass. 

\subsection{The dependence on the planetary mass}\label{sec:result2}

\subsubsection{The positions of inflow and outflow}\label{sec:result3}
The height of the starting point of gas falling increased with the planetary mass, due to the planet's gravity becoming stronger. The width of the horseshoe streamline also widened with increasing planetary mass, and this tendency was consistent with the result of \cite{Fung:2015}. 

When we analysed the velocity of the flow field around the planets, especially the estimation of the speed of outflow (described later in subsection \ref{sec:outflow}), it was necessary to judge where the gas chiefly flowed in and out. Figure \ref{fig:massflux} shows the azimuthally averaged mass flux, $\langle\rho v_{r}\rangle_{\phi}$, as a function of the altitude, $z$. The altitude is changed along with the characteristic spherical surface: the Bondi sphere (blue), the Hill sphere (orange), twice the size of the Bondi sphere (green), and twice the size of the Hill sphere (red), respectively. In \Figref{fig:massflux}, when the azimuthally averaged mass flux has a positive or negative value, it means that gas exits or enters a certain region corresponding to each solid line. Inflow and outflow are balanced in each region where $\langle\rho v_{r}\rangle_{\phi}=0$. As shown in \Figref{fig:massflux}, the maximum and the minimum value of the azimuthally averaged mass flux decreases with increasing radius. Azimuthally averaged mass flux has the maximum value at the mid-plane of the disc and the minimum value at the zenith of the Bondi (for \texttt{m001, m005, m01, m05}) or Hill region (for \texttt{m1, m2}). Though \Figref{fig:meridian01} shows that gas flows in at a considerably high altitude (about twice as high as the Bondi radius of the planet), the quantitative analysis of the azimuthally averaged mass flux tells us that gas mainly flows in and out of whichever is smaller: the Bondi or Hill radius. 

In \texttt{m001}, \texttt{m005}, and \texttt{m01} runs, we set the size of the outer boundary to be smaller than the disc scale height. As shown in the enlarged view in the lower right in Figs. \ref{fig:massflux}a-c, the dominant mass flux occurs at the position of the Bondi radius (blue solid line). In the outer region of the Bondi sphere, mass fluxes are almost zero (orange and green solid lines). Therefore, when the planetary mass was small, the materials that existed outside the Bondi radius hardly accreted and do not contribute to the changes of the flow field. We confirmed these arguments were valid through the results of \texttt{m001-extendD, m005-extendD} and \texttt{m01-extendD}  runs whose domain sizes were taken to be larger than \texttt{m001}, \texttt{m005}, and \texttt{m01} runs (see Appendox \ref{sec:appendix} for details). These results justify our assumption on the domain sizes.


\subsubsection{The outflow speed}\label{sec:result4}
The speed of outflow near the Bondi or Hill radius at the mid-plane of the disc increased with planetary mass (\Figref{fig:outflowlog}). This figure shows the differences of the outflow speed between different planetary masses at $t=t_{\rm end}$. Colour contour represents the radial velocity of the flow field in logarithmic scale. The topologies of the flow field are slightly different in each panel Figs. \ref{fig:outflowlog}a-f, especially in the vicinity of the planet, but each panel has an equal amount of three types of streamlines: the Keplerian shear streamlines, the horseshoe streamlines, and the atmospheric recycling streamlines (we have already discussed this universal nature of the flow field in subsection \ref{sec:result1}). The flow field seems to have reached the steady state. In all simulations, the maximum and the minimum values of the radial velocity were hardly changed over time in the late stage of the time evolution. The time until the flow field reaches the steady state was found to be longer as the mass of the planet was larger. Discussion as to whether the flow field has reached the steady state becomes important in deriving the analytic solution of the outflow in subsection \ref{sec:outflow}.

In this study, we defined the outflow region as the section where the radial velocity was dominant in the flow field near the planet compared to the azimuthal velocity; that is, a condition of $v_{r}>v_{\phi}$ was set for the definition of the outflow. Figure \ref{fig:velrp} shows where the effective outflow emerges. The solid lines are the azimuthally averaged radial and azimuthal velocity of gas, $\langle v_{r}\rangle_{\phi}$ and $\langle v_{\phi}\rangle_{\phi}$, as a function of the radius, $r$. The dashed lines represent the location of either the Bondi or Hill radius of the planet. The left panel shows the result of the analysis from the \texttt{m005} run at $t=30$. In the left panel, the radial velocity exceeds the azimuthal velocity near the Bondi radius. On the other hand, in the right panel which shows the result from the \texttt{m2} run at $t=100$, the radial velocity surpasses the azimuthal velocity near the Hill radius of the planet. We also analysed other results of simulations and found that the outflow emerged the smaller radius out of the Bondi and Hill radii. These results are consistent with the analysis of the results of the azimuthally averaged mass flux, which shows where the dominant gas outflow was located (\Figref{fig:massflux}). 

The effective outflow speed has a maximum value at $\phi\sim3\pi/4,\ 7\pi/4$ (\Figref{fig:vrmax}). Inflow or outflow occurs where $v_{r}$ is positive or negative. In the 2D simulation, it is expected that the inflow and the outflow speed will be the same in the mid-plane \cite[]{Ormel:2015a}. Figure \ref{fig:vrmax} shows, however, the maximum and the minimum value of the radial velocity are different.

The dependency of the outflow speed on the planetary mass is presented in \Figref{fig:outflow}. The cross symbols represent the results of the analysis of the simulations at $t=t_{\rm end}$. We plot the maximum value of the effective outflow speed at the smaller radius out of the Bondi and Hill radii on the mid-plane of the disc obtained from the analysis of the results (\Figref{fig:vrmax}). The solid lines coloured blue and orange correspond to the analytic approximate solution of the outflow speed (see the next subsection \ref{sec:outflow}). When the dimensionless planetary mass is smaller than the unity, it seems that the outflow speed increases in proportion to the first power of the planetary mass. However, once the dimensionless planetary mass exceeds the unity, the power-law index in \Figref{fig:outflow} becomes less than unity. How is the outflow speed actually expressed as a function of the planetary mass? Later in subsection \ref{sec:outflow}, we conducted an analytical derivation of the outflow speed.

\subsection{The analytical estimate of outflow speed}\label{sec:outflow}
The analytical solution of the outflow speed was derived from Bernoulli's theorem. Assuming $\bm{u}$ is the gas velocity in the rotating frame, under the isothermal and inviscid condition, Bernoulli's function is described as
\begin{align}
B=\frac{u^{2}}{2}+c_{\rm s}^{2}\ln\rho+\Phi_{\rm eff},
\end{align}
where $\Phi_{\rm eff}$ is the effective potential expressed by
\begin{align}
\Phi_{\rm eff}=-\frac{GM_{\ast}}{\sqrt{r_{\rm \ast}^{2}+z^{2}}}-\frac{GM_{\rm p}}{r}-\frac{r_{\ast}^{2}\Omega^{2}}{2},\label{eq:effective}
\end{align}
where $r_{\ast}$ is the distance from the centre of the central star. The first to the third terms in the RHS of \Equref{eq:effective} correspond to the star's gravitational potential, the planet's gravitational potential, and the centrifugal potential, respectively. When this effective potential is linearly approximated with $r_{\ast}=a+x$ and $|x|,\ |z|\ll a$ , Bernoulli's function can be rewritten as
\begin{align}
B=\frac{u^{2}}{2}+c_{\rm s}^{2}\ln\rho-\frac{\Omega^{2}}{2}\left(3\left(x^{2}+a^{2}\right)-z^{2}\right)-\frac{GM_{\rm p}}{\sqrt{x^{2}+y^{2}+z^{2}}}.
\end{align}

We picked up a streamline of the recycling flow (\Figref{fig:3Dsinglestreamline}). We set two points on the streamline, $P_{1}=(0,0,z_{1})$ and $P_{2}=\left(x_{2},y_{2},0\right)$. These two points correspond to where gas flows in and out. Assuming that inflow has the velocity $\bm{u}=\bm{u}_{\rm in}$ at $P_{1}$ and outflow also has the velocity $\bm{u}=\bm{u}_{\rm out}$ at $P_{2}$.  Since at each point of $P_{1}$ and $P_{2}$ Bernoulli's function has the same value in the steady state, outflow speed can be described as
\tiny
\begin{align}
|u_{\rm out}|=\sqrt{|u_{\rm in}|^{2}+2c^{2}_{\rm s}\ln\frac{\rho_{\infty}(z_{1})}{\rho_{\infty}(0)}+3x_{2}^{2}\Omega^{2}+z_{1}^{2}\Omega^{2}+2GM_{\rm p}\left(\frac{1}{\sqrt{x_{2}^{2}+y_{2}^{2}}}-\frac{1}{z_{1}}\right)}.\label{eq:outflowspeed}
\end{align}
\normalsize
The second and fourth term in the RHS of \Equref{eq:outflowspeed} would be cancelled because the stellar gravitational potential energy balances $\ln\rho$ in the hydrostatic equilibrium. 

We defined $R_{\rm in}=z_{1}$ and $R_{\rm out}=\sqrt{x_{2}^{2}+y_{2}^{2}}$ for convenience, which meant the distance from the centre of the planet where gas flowed in and out. Since we found that gas chiefly flowed in and out from the smaller of the Bondi and Hill regions--that is, the distance of the inflow and outflow point from the centre of the planet was the same--$R_{\rm in}=R_{\rm out}$. Therefore, the fifth term is also eliminated. In this case, \Equref{eq:outflowspeed} shows the relation $|u_{\rm out}|>|u_{\rm in}|$ and gives the upper limit of the outflow speed. From the analysis of the kinetic energy of gas flow in our simulations, we found the contribution of $|u_{\rm in}|^{2}$ was negligible at $P_{1}$. Consequently, the tidal potential term determines the outflow speed $u_{\rm out}$ in \Equref{eq:outflowspeed}.


We defined \textit{the Bondi} and \textit{the Hill regime} depending on the dimensionless planetary mass. The boundary was located at $m=0.58$ where the size of the Bondi sphere of the planet exceeds the size of the Hill sphere. 
In \textit{the Bondi regime}, $m<0.58$, since the effective outflow leaves the Bondi sphere near the mid-plane of the disc and has the maximum value at $x_{2}\approx R_{\rm Bondi}/\sqrt{2}$, \Equref{eq:outflowspeed} gives
\begin{align}
|u_{\rm out}|_{\rm Bondi}\simeq\sqrt{\frac{3}{2}}mc_{\rm s}.\label{eq:bondiregime}
\end{align}
In \textit{the Hill regime}, $m>0.58$, from the similar procedure we set $x_{2}\approx R_{\rm Hill}/\sqrt{2}$ and using \Equref{eq:outflowspeed} we obtain,
\begin{align}
|u_{\rm out}|_{\rm Hill}\simeq\sqrt{\frac{3}{2}}\left(\frac{m}{3}\right)^{1/3}c_{\rm s}.\label{eq:hillregime}
\end{align}
Our analytic approximate solutions of the outflow speed are plotted in \Figref{fig:outflow}. The outflow speed increases with the dimensionless planetary mass. The power-law index depends on the regimes. It is found that the analytical estimate reproduced the results of our simulations. Since Bernoulli's theorem is applicable only to the steady flow, the agreement is consistent with the inference that our simulations at $t=t_{\rm end}$ have reached the steady state. We note that the expression of $u_{\rm out}$ (\cref{eq:bondiregime,eq:hillregime}) represents the outflow speed not at an arbitrary point along the recycling streamline, but at a specific point where $x=-y=\min\left(R_{\rm Bondi},R_{\rm Hill}\right)/\sqrt{2}$, because we assumed $R_{\rm in}=R_{\rm out}$.

The outflow speed was similar to that of the local Keplerian shear, especially in \textit{the Bondi regime}. However, one slight difference is that the local Keplerian shear velocity is expressed by $v_{\rm shear}=-3/2x\Omega,$ where $x$ is the position in the radial direction and $\Omega$ is the Keplerian frequency. By substituting $x=R_{\rm Bondi}/\sqrt{2}$ which corresponds to the outflow point, we obtained $v_{\rm shear}=3\sqrt{2}/4mc_{\rm s}$. Therefore, the outflow speed in \textit{the Bondi regime} was slightly faster than the local Keplerian shear velocity. 

We measured the outflow speed at $r=\min(R_{\rm Bondi}, R_{\rm Hill})$ because this scale gives the size of the region where the flow is largely influenced by the gravity of the planet. However, the outflow extends beyond $r=\min(R_{\rm Bondi}, R_{\rm Hill})$ and the speed can be estimated by \Equref{eq:outflowspeed} (Appendix \ref{sec:appendix2}).

\section{Discussion} \label{sec:discussion}
\subsection{Application to the non-isothermal simulations}
Our method of the estimate of outflow speed is expected to be applicable to non-isothermal simulations. Under isothermal conditions, previous studies and our results have shown that gas enters at high latitudes of the Bondi or Hill sphere of the planet and leaves it through the mid-plane region of the disc \cite[]{Ormel:2015b,Fung:2015,Kurokawa:2018}. However, non-isothermal simulations have suggested  different trends. A region emerges where a part of the gas is bound around the core of the planet. The 3D radiation-hydrodynamical simulations on the global frame have identified the interface which divides the materials bound by the planet and the unbound ones lie at $\sim0.4R_{\rm Bondi}$ in their 5--10 au runs for $m=0.08$--$0.25$ \cite[]{DAngelo:2013}. By conducting 3D radiation-hydrodynamical inviscid simulations for $m=0.04,0.38,0.75$, and 1.9 planets, \cite{Cimerman:2017} have indicated there are no out-spiralling streamlines in the mid-plane corresponding to the outflow streamline as shown in Fig.8 of \cite{Ormel:2015b} or \Figref{fig:3Dsinglestreamline}. Instead, streamlines are circulating close to the planet. The opacity of the disc also affects the structure of the envelope \cite[]{Lambrechts:2017}. Their three-dimensional radiation-hydrodynamical simulations on the global frame with the opacity $\kappa=0.01\ {\rm cm^{2}/g}$ have found a three-layer structure inside the envelope which consists of the advection layer in the outer layer, the radiative layer in the middle layer, and the convection layer in the inner layer. These differences between isothermal and non-isothermal calculation results come from the buoyancy barrier in the envelope \cite[]{Kurokawa:2018}. They have performed two types of three-dimensional hydrodynamical simulations on the local grid: isothermal and non-isothermal cases. In the case of non-isothermal simulations, the inflow is prevented from reaching the deep part of the envelope because buoyant force suppresses intrusion of high-entropy gas into the low-entropy atmosphere when the atmosphere starts cooling. 

In the non-isothermal simulations, though the atmospheric recycling has only been observed outside the isolated inner envelope, it has not completely disappeared  \cite[]{Kurokawa:2018}. We derived the analytical solution of the outflow speed from Bernoulli's theorem along an atmospheric recycling streamline under the isothermal condition. In the non-isothermal case, the form of Bernoulli's function has to be changed according to the conditions. As long as the atmospheric recycling has not completely disappeared, it is expected that our method can be applied even for non-isothermal simulations. However, there is a possibility that the gas inflow and outflow points may change, and further studies on the current topic are therefore required.

\subsection{Comparison to the analytic solution of \cite{Fung:2015}}
The analytical solution shown in subsection \ref{sec:outflow} differs from that of \cite{Fung:2015}. In their study, the outflow speed has been given as
\begin{align}
|u_{\rm out}|=\sqrt{u_{\rm in}^{2}+2GM_{\rm p}\left(\frac{1}{R_{\rm out}}-\frac{1}{\sqrt{R_{\rm out}^{2}+R^{2}_{\rm in}}}\right)},
\end{align}
and approximately expressed by
\begin{align}
|u_{\rm out}|\simeq
\begin{cases}
\sqrt{2\sqrt{m}}c_{\rm s}, \ &(\text{for}\ m\ll1) \\
\sqrt{\dfrac{3}{8}}c_{\rm s}, \ &(\text{for}\ m\gg1)
\end{cases}\label{eq:fungoutflow}
\end{align}
where the contribution of $u_{\rm in}^{2}$ has been neglected. Equation (\ref{eq:fungoutflow}) predicts when the planetary mass is quite small, $m\ll1$, the outflow speed increases with the planetary mass, but after that, it becomes constant as the planetary mass is sufficiently large. The maximum value of the outflow speed is about $u_{\rm out}\simeq0.6c_{\rm s}$. However, these analytic solutions are inconsistent with our simulation results. Our results showed the outflow speed can be slower in the Bondi regime and faster in the Hill regime than their prediction (\Figref{fig:outflow}). The differences between \cref{eq:bondiregime,eq:hillregime} and \Equref{eq:fungoutflow} came from the following reasons.

Utilisation of Bernoulli's theorem: In their study, Bernoulli's theorem has been applied along with a certain horseshoe streamline. On the other hand, we applied Bernoulli's theorem for an atmospheric recycling streamline by assuming that gas entered at the zenith of the Bondi or Hill sphere. Whereas they have assumed inflow and outflow only have $x$- and $y$-component, $\bm{u}=(u_{x},3/2x\Omega,0)$ where $3/2x\Omega$ is derived from the local Keplerian shear, we did not assume the components of the inflow and outflow (see Appendix of \cite{Fung:2015} for details).

Configurations of inflow and outflow points: Assumption on the altitude where the gas starts descending was also different. They have considered the possibility that gas starts falling from the disc scale height, $R_{\rm in}\approx H$. From the above-mentioned discussion (subsection \ref{sec:result3}), we found gas mainly entered or exited whichever Bondi or Hill region of the planet was smaller. In this study we assumed the height of the falling point was the Bondi or Hill radius, $R_{\rm in}=\min(R_{\rm Bondi},\ R_{\rm Hill})$. Though they have also assumed the distance of the outflow point from the centre of the planet, $R_{\rm out}$, scales with the half-width of the horseshoe orbit for the low mass limit, $m\ll1$, and scales with the Hill radius for the high mass limit, $m\gg1$, we assumed the outflow point was located at the Bondi or Hill radius, $R_{\rm out}=\min(R_{\rm Bondi},R_{\rm Hill})$. 

Therefore, our analytic approximate solution of the outflow speed was different from that of theirs\footnotemark[1]. Since the inflow and outflow points are different from their settings, it is difficult to compare our results to theirs any further. \\ \footnotetext[1]{We note that they have measured the radial outward flows, which flow not from $R_{\rm Bondi}$ but $0.5R_{\rm Bondi}$ in the mid-plane region. Their numerical result has shown the outflow has a speed of $\sim0.2c_{\rm s}$. If we assume $m=0.56$ and $x_{2}\approx 0.5\times R_{\rm Bondi}/\sqrt{2}$ which corresponds to the value used in \cite{Fung:2015}, \Equref{eq:bondiregime} gives us $|u_{\rm out}|\sim0.34c_{\rm s}$. This prediction is consistent with their result.}

\subsection{Implications for the formation of super-Earths via pebble accretion}\label{sec:relativevelocity}
We propose that the ubiquity of super-Earths may be explained by their late-stage formation due to the outflow barrier: the recycling outflow prevents dust and pebbles from accreting onto their proto-cores. In the pebble accretion theory \cite[]{Ormel:2010,Lambrechts:2012}, proto-cores can grow to a mass heavy enough to carve a gap in the pebble disc, which is given by \citep{Lambrechts:2014},
\begin{align}
M_{\rm iso}^{\rm pebble}\sim20\left(\frac{H/a}{0.05}\right)^{3}M_{\oplus}.
\end{align}
This mass is large enough to allow disc gas to accrete in a runaway fashion within the lifetime of a protoplanetary disc \citep{Lee:2014}. 

Simulations in this study and previous ones showed that a planet embedded in a protoplanetary disc induces outflow in the mid-plane region. The particle scale height is given by \citep{Youdin:2007},
\begin{align}
H_{\rm d}=H\left(1+\frac{{\rm St}}{\alpha}\frac{1+2{\rm St}}{1+{\rm St}}\right)^{-1/2},
\end{align}
where St is the dimensionless Stokes number and $\alpha$ is the viscosity parameter \citep{Shakura:1973}. Assuming ${\rm St}\sim10^{-2}\text{--}10^{-1}$ and $\alpha\sim10^{-4}$ to $10^{-2}$ gave $H_{\rm d}/H\sim10^{-2}$ to $10^{0}$. Because the vertical scale of the outflow is estimated to be a few tens of percent of the Bondi radius (\Figref{fig:meridian01}), planets having the dimensionless mass $m=R_{\rm Bondi}/H>10^{-2}$ to $10^{0}$ have the potential to prevent pebbles from accreting onto them. In 2D cases, the flow field around a planet has been shown to influence the accretion rate of particles \cite[]{Ormel:2013}. 

Three-dimensional adiabatic hydrodynamical simulations of gas and particle dynamics has shown that particles that entered the Hill sphere of the planet later exited it on the outer-trailing horseshoe flow \cite[]{Popovas:2018}. They have reported that the dominant inflow of the particles is relevant to the inner-trailing and outer-leading horseshoe flow, and dominant outflow relates to the outer-trailing and inner-leading horseshoe flow. In the Bondi region, the larger particles ($>0.1$ cm) are rapidly accreted onto the planet whose mass is $m=0.011,0.037$ and 0.07. However, since the outer-trailing horseshoe flows are strong enough to carry out the smaller particles ($<0.1$ cm), they do not accrete onto the planet.

We compared the outflow speed to the terminal velocity of particles within the Bondi or Hill radius in order to discuss the influence of the outflow on the core growth. Given the force balance between the gas drag and the planet's gravity acting on the particle,
\begin{align}
\frac{\Delta v}{t_{\rm stop}}\sim\frac{GM_{\rm p}}{r^{2}},\label{eq:particlevelocity}
\end{align}
where $\Delta v$ is the terminal speed of the particle relative to the gas, $t_{\rm stop}$ is the stopping time expressed by $t_{\rm stop}={\rm St}/\Omega$. 

In the Bondi regime, we substituted $r=R_{\rm Bondi}$ into \Equref{eq:particlevelocity} and obtained
\begin{align}
\Delta v\sim\frac{GM_{\rm p}}{R^{2}_{\rm Bondi}}\frac{\rm St}{\Omega}=m^{-1}{\rm St}\ c_{\rm s}.\label{eq:relativebondi}
\end{align}

In the Hill regime, we obtained
\begin{align}
\Delta v\sim\frac{GM_{\rm p}}{R^{2}_{\rm Hill}}\frac{\rm St}{\Omega}=3^{2/3}m^{1/3}{\rm St}\ c_{\rm s}.\label{eq:relativehill}
\end{align}

We plotted the changes of the relative velocity of the small particles as a function of the dimensionless planetary mass for each Stokes number, ${\rm St}=10^{-3},10^{-2}$, and $10^{-1}$ in \Figref{fig:outflow}. As shown in this figure, the flow field around the planets may affect the accretion of the small particles when the outflow speed exceeds the relative velocity of the particles. The planetary mass having the potential to affect the accretion can be written as $m\gtrsim\sqrt{\rm St}$. The width of the outflow is wider than that of inflow in the mid-plane region near the Bondi or Hill radius (see Figs. \ref{fig:mid-plane} and \ref{fig:outflowlog}). It is expected that the outflow reduces the accretion of solid materials onto the planet if they enter the Bondi or Hill sphere from the outflow regions.

Even if the solid particles enter the Bondi or Hill sphere from the inflow region, the accretion onto the planet may be prevented by the flow inside the sphere. As shown in \Figref{fig:vicinity}, outflow is dominant in the region close to the planet. The solid materials supplied from the inflow window of the Bondi or Hill sphere may be transported outwards by the outflow and are ultimately ejected from the envelope. The dense and hot envelope in the vicinity of the planet may also induce the disruption and vaporisation of solid particles, which would further prevent accretion onto the planet \cite[]{Alibert:2017}, though the inner part of the envelope may be isolated from the recycling flow \cite[]{Kurokawa:2018}. Furthermore, even the particles with a relatively large Stokes number (defined at the Bondi or Hill radius)---for instance ${\rm St}>m^{2}$---may also be affected by the outflow within the Bondi or Hill sphere. Since the Stokes number is inversely proportional to the gas density both in the Epstein and Quadratic regimes \cite[e.g.,][]{Ormel:2010}, the effective Stokes number is considered to decrease inside the envelope where gas density is much higher than that of the background. In such cases, the particles become more susceptible to the outflow barrier.

Our results suggest that the flow in the vicinity of proto-cores would delay the formation of super-Earth cores and, consequently, help them to avoid the runaway gas accretion within the disc's lifetime. We propose a plausible scenario of the formation of super-Earths as follows.
\begin{enumerate}
\item Proto-cores form in the outer region ($\sim1$ au) of the disc under the influence of the flow field. Due to the outflow barrier, the growth of proto-cores may halt when $m\sim\sqrt{\rm St}$.
\item When the growth of the proto-cores halts, they begin to migrate inwards. A plurality of proto-cores are arranged at the inner edge of the disc.
\item Super-Earths are formed by giant impact during disc dispersal. In a short time, until the gas has dissipated, an envelope forms around the super-Earths, which has 1--10\% the mass of it. 
\end{enumerate}

In either case, further studies on the interaction of the planet-induced wind with solid materials are needed to understand the consequences on the formation scenarios of super-Earths.


\section{Conclusions} \label{sec:conclusion}
We investigated gas flows around an embedded planet in a protoplanetary disc, and the dependency of the flow field on the planetary mass. We considered isothermal, inviscid gas flow, and performed a series of three-dimensional  hydrodynamical simulations on a spherical polar grid that had a planet placed at its centre. We summarise our main findings as follows.
\begin{enumerate}
\item The three-dimensional structure of the flow field did not change significantly even if we changed the mass of the planet. Gas entered at high latitudes of the Bondi or Hill sphere and left it through the mid-plane of the disc, which was consistent with previous works \cite[]{Ormel:2015b,Fung:2015,Cimerman:2017,Lambrechts:2017,Kurokawa:2018}. The flow field had three types of streamlines: the Keplerian shear streamlines, the horseshoe streamlines, and the atmospheric recycling streamlines. 
\item Gas flowed in and out substantially from the smaller of the Bondi and Hill regions. Azimuthal averaged mass flux increased with the mass of the planet. It had a maximum peak at the mid-plane of the disc, and had a minimum value at the zenith of the Bondi or Hill sphere. 
\item Outflow speed increased with planetary mass. Under isothermal circumstances, we derived an analytical solution of the outflow speed from Bernoulli's theorem. Our equation predicted that the following relations held for the Bondi regime ($R_{\rm Bondi}<R_{\rm Hill}$): $|u_{\rm out}|=\sqrt{3/2}\ mc_{\rm s}$, for the Hill regime ($R_{\rm Bondi}>R_{\rm Hill}$): $|u_{\rm out}|=\sqrt{3/2}\ (m/3)^{1/3}c_{\rm s}$. These predictions were consistent with the results of numerical simulations. 
\item Comparing these analytic solutions and the relative velocity of the small particles, we estimated the dimensionless planetary mass having the potential to affect the accretion of solid materials as $m\gtrsim\sqrt{\rm St}$. As the mass of the planet increased, the outflow became fast and would start to prevent solid materials from accreting onto the core.
\end{enumerate}
Our results suggested the flow field around a planet had the potential to affect the accretion rate of solid materials. It is possible that the outflow barrier could inhibit the accretion of small particles. This mechanism may delay the growth of the solid core of the planet and may be helpful to explain the formation of super-Earths. 

\begin{acknowledgements}
We thank Athena++ developers: James M. Stone, Kengo Tomida, and Christopher White. The authors are grateful for the constructive feedback from an anonymous referee. This study has greatly benefited from fruitful discussion with Chris W. Ormel, Michiel Lambrechts, and Anders Johansen. HK was supported by JSPS KAKENHI Grant number 16H04073, 17H06457, and 18K13602. SI was supported by JSPS KAKENHI grant 15H02065. Numerical computations were in part carried out on Cray XC30 at Earth-Life Science Institute and at the Center for Computational Astrophysics, National Astronomical Observatory of Japan. This research was supported by a grant from the Hayakawa Satio Fund awarded by the Astronomical Society of Japan.
\end{acknowledgements}



\appendix
\def\theequation{A.\arabic{equation}}
\def\thefigure{\Alph{section}.\arabic{figure}}

\section{The dependence on computational resolution and domain size} \label{sec:appendix}


 \begin{figure}[!htbp]
 \resizebox{\hsize}{!}
 {\includegraphics{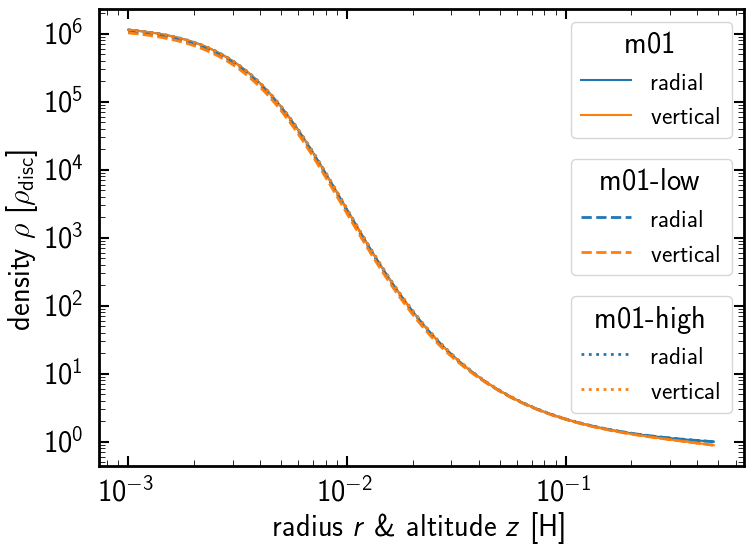}} 
 \caption{The gas density in the radial and vertical direction as a function of the radius, $r$ and altitude, $z$. These results are obtained from \texttt{m01}, \texttt{m01-low}, and \texttt{m01-high} runs at $t=30$. The solid, dashed, and dotted lines correspond to the radial (blue) and vertical (orange) gas density around the planet obtained from the fiducial, low, and high resolution simulations, respectively. In the radial direction, we plotted the azimuthally averaged gas density. In the vertical direction, we plotted the gas density along $z-$axis.}
\label{fig:density}
\end{figure}

 \begin{figure}[!htbp]
 \resizebox{\hsize}{!}
 {\includegraphics{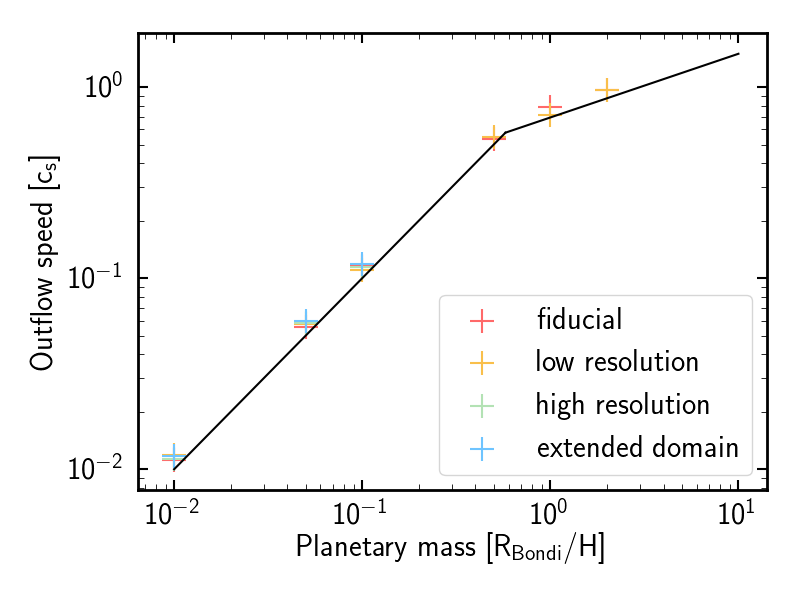}} 
 \caption{The maximum value of the outflow speed at the Bondi (for the planets having $m=0.01, 0.05, 0.1$, and $0.5$) or Hill radius (for the planets having $m=1.0$ and $2.0$) on the mid-plane of the disc obtained from the fiducial (red), low resolution (yellow), high resolution (green), and extended domain (blue) simulations at $t=t_{\rm end}$. The black solid line corresponds to the analytical solution of the outflow speed derived from the analytical solution of the outflow speed derived from \cref{eq:bondiregime,eq:hillregime}.}
\label{fig:outflowlow}
\end{figure}

To investigate the effect of numerical configuration on the main result and confirm the numerical convergence, we performed simulations with the extended domain size and with the lower or higher resolution ($[\log r,\theta,\phi]=[100,50,100]$ or $[160,80,160]$). All of our simulations are listed in \Tabref{tab:simulation}.

As shown in \Figref{fig:density}, the gas density obtained by fiducial (solid lines), low (dashed lines), and high (dotted lines) resolution simulations agree with each other.

The maximum outflow speed at the mid-plane of the disc also matched. Figure \ref{fig:outflowlow} shows the results obtained by a series of fiducial, low resolution, high resolution, and extended domain simulations. Although there are slight differences in the results of the three simulations, the outflow speed obtained from simulations with different resolutions or domain sizes are also consistent with our analytic solution plotted by the black solid line.

From these results, we concluded that all of our hydrodynamical simulations reached numerical convergence and our choice of the domain size does not affect the main results.

\section{The application of analytic solution}\label{sec:appendix2}
 \begin{figure}[!htbp]
 \resizebox{\hsize}{!}
 {\includegraphics{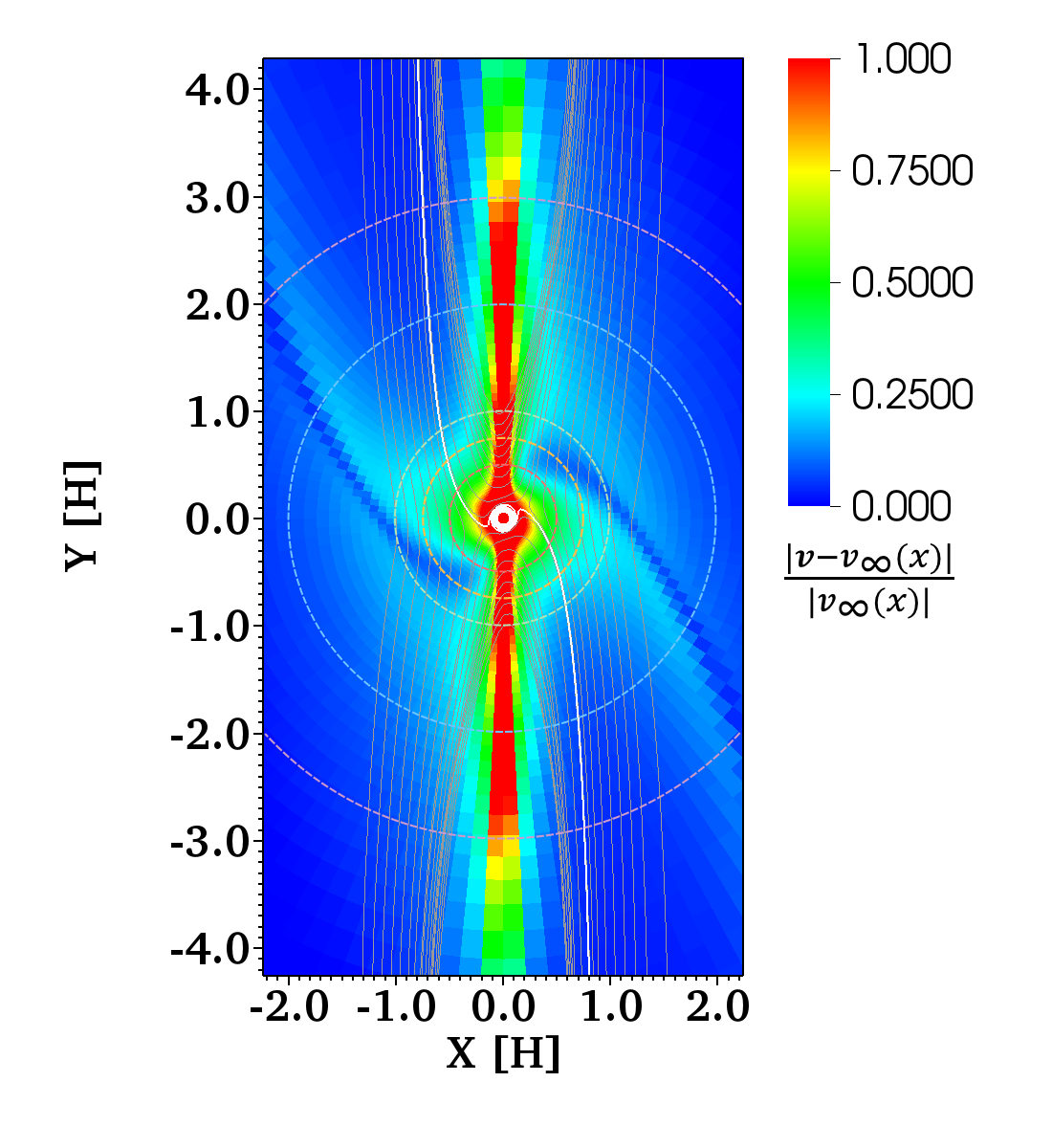}} 
 \caption{The colour contour shows the difference of flow speed from the Keplerian shear obtained by \texttt{m05} run at $t=t_{\rm end}$. We plotted $|\bm{v}-\bm{v}_{\infty}(x)|/|\bm{v}_{\infty}(x)|$, where $\bm{v}$ is the gas velocity and $\bm{v}_{\infty}(x)$ is the Keplerian shear expressed by \Equref{eq:Keplerianshear}. The contour shows that the spherical region around the planet is highly influenced by the gravity. We note that the red region along $x=0$ is not the outflow but the horseshoe flow. We also note that the colour contour is saturated in the region colored with red. The gray solid lines correspond to the specific gas streamlines. We highlighted the atmospheric recycling streamline in the thick white solid line. The red, orange, green, blue, and purple dashed lines are the circles of radius $R_{\rm Bondi}$, $1.5R_{\rm Bondi}$, 1\ $[H]$, 2\ $[H]$, and 3\ $[H]$.}
\label{fig:streamline_long}
\end{figure}

 \begin{figure}[!htbp]
 \resizebox{\hsize}{!}
 {\includegraphics{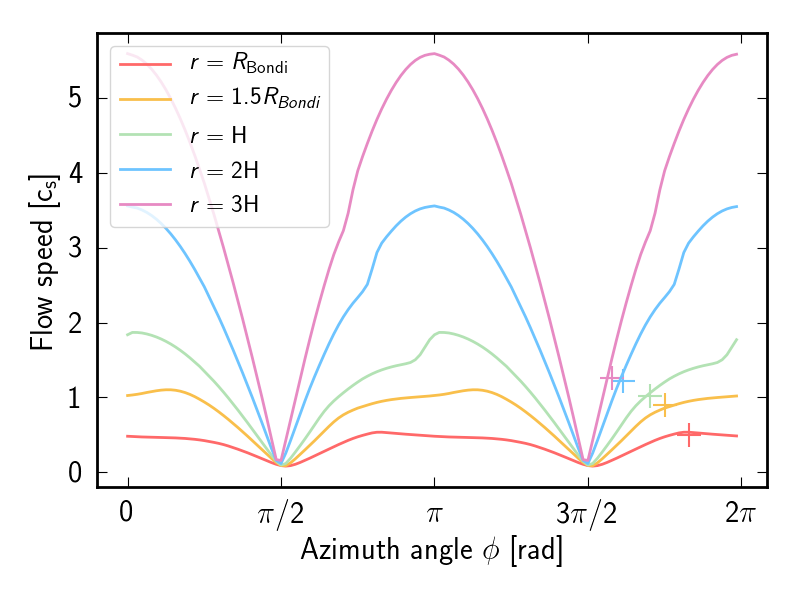}} 
 \caption{The flow speed obtained by \texttt{m05} run as a function of azimuth angle at different $r$ and $t=t_{\rm end}$. The cross symbols coloured by red, orange, green, blue, and purple correspond to the flow speed estimated by \Equref{eq:outflowspeed}, $|u_{\rm out}|\simeq\sqrt{3}\Omega x$, at the intersections of the circles coloured by red, orange, green, blue, and purple with the atmospheric recycling streamline in the forth quadrant plotted in \Figref{fig:streamline_long}.}
\label{fig:vr_curves}
\end{figure}
Because the Bernoulli's principle is valid along a streamline in the steady state, the outflow speed beyond $r=\min(R_{\rm Bondi}, R_{\rm Hill})$ can be estimated from \Equref{eq:outflowspeed}.

Figure \ref{fig:streamline_long} shows the difference of flow speed from the unperturbed Keplerian shear. In this study, we measured the speed of planet-induced outflow at $r=\min(R_{\rm Bondi}, R_{\rm Hill})$ because this scale gives the extent of the flow influenced by the planet gravity. The outflow extends up to 2--3 times $r=\min(R_{\rm Bondi}, R_{\rm Hill})$ and merges with the shear flow.

Figure \ref{fig:vr_curves} shows the flow speed measured at different $r$. Here we assumed that the gravity and density terms canceled and that $|u_{\rm in}|^{2}$ was negligible. The results confirm that the outflow accelerates beyond $x=\min(R_{\rm Bondi}, R_{\rm Hill})/\sqrt{2}$ following \Equref{eq:outflowspeed}.

\end{document}